\documentclass[sigconf, screen]{acmart}
\AtBeginDocument{%
  }

\usepackage{multirow}
\usepackage{tabularx}
\usepackage{ragged2e}
\usepackage{xspace}
\copyrightyear{2026}
\acmYear{2026}
\setcopyright{cc}
\setcctype{by-nc-nd}
\acmConference[CHI '26]{Proceedings of the 2026 CHI Conference on Human Factors in Computing Systems}{April 13--17, 2026}{Barcelona, Spain}
\acmBooktitle{Proceedings of the 2026 CHI Conference on Human Factors in Computing Systems (CHI '26), April 13--17, 2026, Barcelona, Spain}
\acmDOI{10.1145/3772318.3790568}
\acmISBN{979-8-4007-2278-3/2026/04}

\begin{document}

\newcommand{\mxj}[1]{{\color{red} #1}}
\newcommand{\xm}[1]{{\color{blue} #1}}
\newcommand{\haotian}[1]{\textcolor{teal}{[Haotian: #1]}}
\newcommand{\revise}[1]{\textcolor{black}{#1}}
\newcommand{\rerevise}[1]{\textcolor{red}{#1}}

\newcommand{\condone}{\revise{\textit{DuoDrama}}\xspace}
\newcommand{\condtwo}{\revise{Exp-PE}\xspace}
\newcommand{\condthree}{\revise{Eval-NoPE}\xspace}
\newcommand{\condfour}{\revise{Rev-NoPE}\xspace}


\title{\textit{DuoDrama}: Supporting Screenplay Refinement Through LLM-Assisted Human Reflection}




\author{Yuying Tang}
\orcid{0009-0003-1906-2834}
\affiliation{%
  \institution{The Hong Kong University of Science and Technology }
  \city{Hong Kong SAR}
  \country{China}
}
\email{yuying.tang@connect.ust.hk}

\author{Xinyi Chen}
\orcid{0009-0006-3105-6834}
\affiliation{%
  \institution{Zhejiang University}
  \city{Hangzhou}
  \country{China}
}
\email{cadrenaline@zju.edu.cn}

\author{Haotian Li}
\authornote{Haotian Li is the corresponding author.}
\orcid{0000-0001-9547-3449}
\affiliation{%
  \institution{Microsoft Research Asia}
  \city{Beijing}
  \country{China}
}
\email{haotian.li@microsoft.com}

\author{Xing Xie}
\orcid{0009-0009-3257-3077}
\affiliation{%
  \institution{Microsoft Research Asia}
  \city{Beijing}
  \country{China}
}
\email{xingx@microsoft.com}

\author{Xiaojuan Ma}
\orcid{0000-0002-9847-7784}
\affiliation{%
  \institution{The Hong Kong University of Science and Technology}
  \city{Hong Kong SAR}
  \country{China}
}
\email{mxj@cse.ust.hk}

\author{Huamin Qu}
\orcid{0000-0002-3344-9694}
\affiliation{%
  \institution{The Hong Kong University of Science and Technology}
  \city{Hong Kong SAR}
  \country{China}
}
\email{huamin@cse.ust.hk}


\begin{abstract}
  AI has been increasingly integrated into screenwriting practice. In refinement, screenwriters expect AI to provide feedback that supports reflection across the internal perspective of characters and the external perspective of the overall story. However, existing AI tools cannot sufficiently coordinate the two perspectives to meet screenwriters’ needs. To address this gap, we present \textit{DuoDrama}, an AI system that generates feedback to assist screenwriters' reflection in refinement. To enable \textit{DuoDrama}, based on performance theories and a formative study with nine professional screenwriters, we design the \textit{Experience-Grounded Feedback Generation Workflow for Human Reflection} (\textit{ExReflect}). In \textit{ExReflect}, an AI agent adopts an \textit{experience role} to generate experience and then shifts to an \textit{evaluation role} to generate feedback based on the experience. A study with fourteen professional screenwriters shows that \textit{DuoDrama} improves feedback quality and alignment and enhances the effectiveness, depth, and richness of reflection. We conclude by discussing broader implications and future directions.
\end{abstract}




\maketitle

\section{Introduction}

\revise{AI has been integrated into screenwriting practice for audiovisual production (e.g., films and dramas)~\cite{mirowski2023co, tang2025understanding, 10.1145/3716135}, assisting screenwriters in generating materials during creative processes. In the refinement stage, however, screenwriters do not only seek AI-generated content but also expect AI to provide feedback that supports their reflection on scene logic, character motivation, and authorial intent~\cite{tang2025understanding}. Existing research shows that reflection is central to screenplay refinement, a stage widely regarded as decisive for shaping the final screenplay~\cite{epps2016screenwriting, gulino2024screenwriting}. Unlike drafting, which focuses on producing material, refinement requires interpreting the written draft and examining its coherence and narrative function~\cite{batty2025interrogating, batty2018script}. In particular, effective refinement depends on coordinating two complementary perspectives of reflection~\cite{senje2017formatting, mckee2005story, thomas2013script}: an internal perspective on what a character thinks or feels and whether each line conveys a credible intention or shift, and an external perspective assessing how that moment influences pacing, causality, and thematic development across the broader story~\cite{mckee2005story}. Maintaining and coordinating both perspectives is difficult in practice~\cite{batty2025interrogating}, yet no existing tool offers support that helps screenwriters coordinate these two perspectives of reflection during refinement.} 

\revise{In the broader field, prior AI systems supporting reflection in creative writing mainly rely on role-playing to deliver feedback~\cite{10.1145/3613904.3642406, grigis2024playwriting, 10.1145/3613904.3642105, rashkin2025help}.
Earlier studies show that AI can adopt persona-based roles and help writers view their work from new perspectives~\cite{10.1145/3613904.3642406, grigis2024playwriting}. Such feedback can encourage writers to reflect on their choices and consider revisions~\cite{10.1145/3613904.3642105, rashkin2025help}. However, existing systems typically generate feedback from a single perspective, such as a role's personality or a character’s appearance~\cite{10.1145/3613904.3642105, rashkin2025help}. 
While this approach is helpful during drafting, such as when developing character designs, it does not meet the reflection needs of refinement, where screenwriters need to coordinate both internal and external perspectives.}
\revise{Recent multi-agent role-playing systems allow AI to generate feedback from multiple perspectives~\cite{zhang2025revtogether, talaei2025storysage}, suggesting its potential for exposing users to diverse perspectives at the same time. However, agents in such systems tend to operate independently, so their feedback cannot be coordinated into a coherent and contextually aligned assessment~\cite{zhang2025revtogether, robertson2002feedback}.
As a result, users still need to reconcile fragmented or misaligned feedback, which is challenging in creative writing contexts~\cite{zhang2025friction, tang2025understanding}. }

\revise{To address this limitation, we introduce \textit{DuoDrama}, a system designed to support coordinated shifting across internal and external perspectives to assist reflection in screenwriters' refinement. 
\textit{DuoDrama} is powered by a new \textit{Experience-Grounded Feedback Generation Workflow for Human Reflection} (\textit{ExReflect}). In our system, we apply \textit{ExReflect} to enable an agent to perform sequential role-playing: it first adopts a screenplay character as its internal \textit{experience role} to generate inner thoughts as personal experience, and then shifts to the actor portraying that character as its external \textit{evaluation role} to generate feedback, using the personal experience as additional interaction context.
The \textit{ExReflect} workflow draws on performance theory: Stanislavski’s immersive embodiment supports reasoning from within a character’s internal perspective~\cite{moore1984stanislavski}, and Brecht’s critical distancing supports assessment from an external perspective~\cite{brecht2013short}. Prior studies indicate that combining these perspectives enables internally situated understanding alongside externally interpretive evaluation~\cite{bentley1964stanislavski, mumford1995brecht}. Based on our formative study with nine professional screenwriters, we summarized four design goals and derived five dimensions that screenwriters need support to reflect on during refinement. Building on these insights, we further extracted the detailed requirements for \textit{ExReflect} in this context.
We then implemented multiple instances of \textit{ExReflect} using a multi-agent architecture in \textit{DuoDrama}, as screenplays typically involve multiple characters whose perspectives should be treated independently when generating feedback for reflection. A user study with fourteen screenwriters showed that \textit{DuoDrama} strengthened screenwriters' reflection by providing high-quality, contextually aligned, and well-timed feedback, resulting in improved effectiveness, depth, and richness of reflection. In summary, this work makes three main contributions:}


\begin{itemize}
\item \revise{
Identifying design goals for AI systems to provide feedback for reflection in screenplay refinement through a formative study. The design goals characterize the perspective, timing, content, and style of AI feedback. }
\item \revise{Developing \textit{DuoDrama}, a system that employs the performance-theory-inspired workflow \textit{ExReflect} to generate experience-grounded feedback that coordinates both internal and external perspectives, supporting screenwriters' reflection in refinement.}
\item \revise{Demonstrating through a user study with fourteen professional screenwriters that \textit{DuoDrama} improves the quality and alignment of AI-generated feedback and enhances screenwriters’ reflection in refinement.}
\end{itemize}

\section{Related Work}



\revise{Our research builds on the understanding of reflection in screenplay refinement. We first outline the needs and challenges screenwriters face during this process. We then review prior work on AI for screenwriting reflection and identify why existing approaches fall short. Finally, to clarify how and when AI can support reflection in refinement, we examine prior studies on AI-assisted human reflection strategies.}

\subsection{\revise{Reflection in Screenwriting Refinement}}

\revise{Screenwriting studies describe refinement as a significant stage in which screenwriters rewrite the draft they have produced and reinterpret each specific moment in relation to the overall evolving narration~\cite{flanagan2025autoethnographic, epps2016screenwriting, gulino2024screenwriting}. Screenwriters do not simply adjust dialogue or pacing, they further understand what each dialogue in a different scene conveys emotionally, structurally, and thematically, and determine whether it continues to serve the broader intentions in the whole storyline~\cite{epps2016screenwriting, nelmes2008developing, lazarus2025rewriting}. Therefore, prior works describe screenwriting refinement as a reflective process that involves repeatedly coordinating between two perspectives: one grounded in the character’s immediate, internally situated experience and one grounded in the externally overview structure~\cite {senje2017formatting, mckee2005story, thomas2013script}.}

\revise{The internal perspective guides the screenwriter to examine psychological credibility, emotional transitions, and moment-to-moment intention by reasoning from within what the character thinks or feels at that point in the story~\cite{selbo2011constructive, moritz2013scriptwriting, thompson1999storytelling, kallas2017creative}. The external perspective, in contrast, situates that moment in the overview design of the script, asking whether it advances causality, supports pacing, strengthens theme, or maintains continuity across characters and scenes~\cite{gulino2024screenwriting, bordwell2006way, brooks1999metalinear}. These perspectives are difficult to hold in balance during refinement. Screenwriters often default to a single interpretive mode, which limits their ability to consider alternative possibilities or spot issues that emerge only from a contrasting perspective~\cite{batty2025interrogating, batty2018script}. Prior work further shows that deep familiarity with one’s own draft narrows attention, reducing sensitivity to narrative inconsistencies, weak motivations, or unclear emotional cues that would be more visible from another perspective~\cite{gulino2018science, plantinga2018screen, mckee1997substance}. Together, these challenges point to the need for support that can coordinate both perspectives at once to help screenwriters surface issues that are missed during their refinement, thereby better supporting their reflection.}


\subsection{\revise{AI for Screenwriting Reflection}}

\revise{AI has increasingly been used in screenwriting not only for direct content generation but also for providing feedback that supports writers’ reflection~\cite{tang2025understanding}. Screenwriters expect AI to move beyond producing alternative lines or scenes~\cite{chen2024hollmwood, mirowski2023co} and to participate in the creative process through feedback offered from different roles, addressing character motivation, scene logic, and authorial intent, for example, by acting as an actor~\cite{tang2025understanding}.}

\revise{Prior work shows that role-playing has been the primary approach through which AI provides feedback for writers’ reflection, demonstrating that feedback from different personas can help writers view their stories from new perspectives~\cite{10.1145/3613904.3642406, grigis2024playwriting}, trigger revisions~\cite{rashkin2025help}, and stimulate reflection~\cite{10.1145/3613904.3642105}. For example, Qin et al. allow writers to interact directly with AI characters~\cite{10.1145/3613904.3642105}. These embodied reactions provide feedback that reveals characters' emotions and motivations and help writers identify new possibilities for character behavior. However, they often remain confined to local interactions and do not connect the character’s internal perspective to scene pacing, thematic development, or overall narrative coherence. As a result, writers receive only internal, momentary emotional insight without support for linking it to the broader external story structure. Other studies instead allow writers to define evaluator personas that provide critiques from a specific external identity~\cite{10.1145/3613904.3642406, rashkin2025help}. These critiques offer a useful external perspective but lack access to the character’s situated experience. As a result, evaluator personas only provide external perspective reflection and struggle to reason about how emotional or motivational cues arise from the character’s internal perspective, offering limited support for checking whether internal psychology aligns with specific actions or dialogue. Overall, existing role-playing systems tend to emphasize either an internal, character-based perspective or an external, evaluator-based perspective, rather than coordinating both in ways that meet the reflection needs of screenplay refinement.}

\revise{Recent systems introduce multi-agent role-playing to provide feedback from multiple perspectives simultaneously~\cite{zhang2025revtogether, talaei2025storysage}. Having multiple agents review a draft through different roles can expose writers to diverse and sometimes conflicting comments, partially supporting perspective shifting. Yet because these agents operate independently, their feedback is not automatically coordinated. Prior studies on feedback show that unstructured or unintegrated comments are difficult to interpret and may feel misaligned with the writer’s context and intent~\cite{zhang2025revtogether, robertson2002feedback, tang2025understanding}. Writers often need to spend considerable effort filtering low-quality suggestions, reconciling conflicting perspectives, and judging which comments align with their goals~\cite{zhang2025friction, tang2025understanding}. Consequently, current multi-agent approaches still do not support the coordination of two perspectives that screenplay refinement requires, where writers should move fluidly between internal character experience and an external overview structure. To address this gap, we develop a system that not only provides both perspectives but also coordinates them into feedback that is high-quality and contextually aligned with the internal and external perspectives of screenplay refinement.}

\subsection{\revise{AI-Assisted Human Reflection Strategy}}
\revise{To understand how human reflection may be supported in screenplay refinement, we review the forms of AI-assisted reflection in prior work. In defining the type of reflection discussed here, we follow prior work on human reflection in HCI and include studies in which the authors explicitly describe their systems as supporting reflection or report that users showed reflective behavior~\cite{bentvelzen2022revisiting, baumer2014reviewing}. Existing AI systems provide three main types of feedback to assist human reflection: suggestions, critiques, and questions. Suggestions offer direct edits or rewritten alternatives that can resolve issues, but they often limit opportunities for interpretation and deeper reflection because writers receive solutions rather than support for examining their own reasoning~\cite{guo2025pen, zhang2025friction}. Critiques present assessments from an external perspective, yet without grounding in the situated context, they tend to be abstract or generic and do not fully account for the intentions or circumstances of a specific narrative moment~\cite{10.1145/3757566, fischer1993embedding}. Question-based feedback prompts writers to clarify intentions, examine assumptions, and reconsider decisions, and prior work shows that such prompts can facilitate human reflection~\cite{10.1145/3757551, 10.1145/3290605.3300368, kreminski2024intent}. Therefore, question-based feedback is particularly well-suited to the demands of screenwriters' reflection in screenplay refinement, where the screenwriter needs to relate each specific decision to the development of the story as a whole.}

\revise{Beyond the content of feedback for supporting human reflection, prior work also varies in when AI offers support. Some systems provide proactive feedback, where the AI initiates comments without explicit requests~\cite{10.1145/3706598.3713760}, while others use on-demand feedback that users pull when needed~\cite{10.1145/3613904.3642406}. Proactive feedback can surface blind spots that writers may not think to query, but it risks interruption or overload if poorly timed. In contrast, on-demand feedback respects writer control but may cause missed opportunities for reflection when writers are uncertain about what to ask. Additionally, research on reflection timing further distinguishes support that occurs \textit{in-action} (during an activity) from support that occurs \textit{on-action} (afterward)~\cite{10.1145/3635636.3656183}, echoing distinctions between reflection-in-action and reflection-on-action~\cite{munby1989reflection}. In-action feedback can help writers notice issues as they work through a specific moment internally, while on-action feedback can encourage external reflection on structure and theme after a draft segment is complete. Prior studies suggest that each timing has different strengths and weaknesses. Building on these insights, we focus on question-based feedback and consider both in-action and on-action timing in our formative study to further understand when screenwriters prefer AI to support their reflection in refinement.}

\section{The \textit{DuoDrama} System}
\revise{To enable \textit{DuoDrama}, we designed the \textit{Experience-Grounded Feedback Generation Workflow for Human Reflection} (\textit{ExReflect}), a workflow inspired by performance theories, to support screenwriters' reflection in refinement. A formative study with nine professional screenwriters helps us understand the detailed requirements for \textit{ExReflect} in this context, producing four design goals that guided the system design.}


\subsection{ExReflect: An \revise{Experience-Grounded Feedback Generation Workflow for Human Reflection} Informed by Performance Theory}

\revise{We design a workflow that originates from a conceptual formulation but is specified in a way that affords technical implementation (see Sec.~\ref{sec:Implementation}). The workflow, inspired by performance theories, integrates immersive enactment and critical detachment within a single agent, enabling AI to support two-perspective reflection. We further detail its computational instantiation in our system (see Sec.~\ref{sec:pre-processing} \& Sec.~\ref{sec:dual_memory}), including the agent's state transitions and data-flow structures.}

\subsubsection{Background}
We draw inspiration from two performance theories as the principled foundation. In professional acting, these traditions cultivate complementary perspectives. Stanislavski’s system emphasizes immersive and psychologically consistent embodiment, encouraging actors to internalize a character’s motivations and emotions in order to deliver an authentic performance~\cite{moore1984stanislavski, carnicke2020stanislavsky}. In contrast, Brecht’s epic theatre stresses critical distance through the Verfremdungseffekt (alienation effect), deliberately breaking identification to expose the constructed nature of performance and prevent audiences from becoming overly absorbed, thereby fostering critical evaluation~\cite{brecht2013short, steer1968brecht}. Prior studies in performance studies have further suggested that integrating these two traditions can yield richer practices, achieving a balance between deep immersion and critical detachment~\cite{bentley1964stanislavski, mumford1995brecht}. \revise{Building on these insights, we propose \textit{ExReflect}, a workflow that combines immersive enactment with critical analysis, grounding feedback in internal experience while maintaining an external perspective for human reflection.}



\begin{figure*}
 \centering         
\includegraphics[width=1\textwidth]{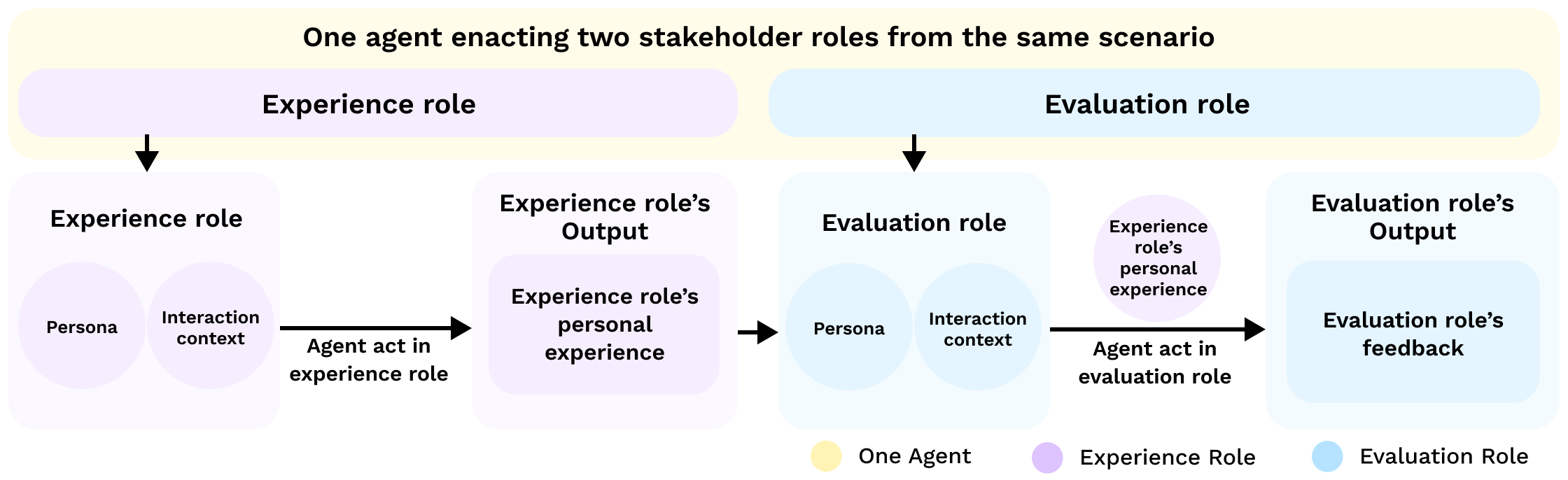} 
\caption{The \textit{Experience-Grounded Feedback Generation Workflow for Human Reflection} (\textit{ExReflect}). A single agent sequentially enacts two stakeholder roles from the same scenario. In the \textit{experience role}, the agent adopts a stakeholder persona and interaction context to generate personal experiences. It then shifts to the \textit{evaluation role}, adopting a different stakeholder perspective to generate feedback based on that personal experience. This design grounds feedback in personal experience while introducing evaluation distance, thereby balancing internal immersion with external critique.
}
 \label{workflow}
\Description{}
 \end{figure*}
 
\subsubsection{\textit{ExReflect} Workflow}  
The \textit{ExReflect} workflow is operationalized within a single agent (Fig.~\ref{workflow}). The user begins by specifying two meaningful stakeholders relevant to the scenario (one \textit{experience role} and one \textit{evaluation role}), each defined by a persona (e.g., background, motivations, traits) and interaction context (e.g., scene, environment, other participants). In the \textit{experience role}, the agent generates internally personal experience, recording emotions, motivations, and situational constraints in line with Stanislavski’s theory of ``living through'' given circumstances. The agent then adopts the \textit{evaluation role} according to Brecht’s alienation effect, shifting to an external stakeholder perspective to assess the scene, using the personal experience generated in the \textit{experience role} as additional interaction context. Guided by these theories, this workflow helps feedback remain grounded in internal context while also introducing an external perspective.

\subsection{Formative Study}
\revise{Before implementing \textit{DuoDrama}, we conducted a formative study with nine professional screenwriters to understand the detailed requirements for \textit{ExReflect} for screenwriters' reflection in refinement.} The study drew inspiration from the Kipling Method~\cite{5681344} and focused on 3W1H: who, what, when, and how. \revise{This method enabled us to systematically elicit screenwriters’ expectations regarding which \textit{experience role} and \textit{evaluation role} an AI agent should assume, what experiences should be generated, when AI feedback should occur, and how AI feedback should be delivered.} Insights from this study were synthesized into four DGs, which directly informed the design of \textit{DuoDrama}. Our study protocol was reviewed and deemed exempt by our university’s Institutional Review Board (IRB).


\subsubsection{Participants and Procedure}
We recruited nine participants with screenwriting experience through snowball sampling~\cite{goodman1961snowball}. Demographic information was self-reported: three identified as male and six as female, with ages ranging from 25 to 39. All participants had prior experience using AI tools in the context of screenwriting. Additional participant details are provided in the supplementary materials. We conducted one-on-one semi-structured interviews guided by the 3W1H:

\begin{itemize}
    \item \textbf{Who:} Roles for \textit{ExReflect}.  
    Participants discussed which \textit{experience role} and \textit{evaluation role} an AI agent should assume, clarifying how different perspectives can be enacted and how they complement each other.  

    \item \textbf{What:} Content and Information Grounded in \textit{ExReflect}.  
    Participants identified which aspects of the simulation and their reflection were most valuable for refinement, including the personal experience the \textit{experience role} should capture and the evaluation metrics the \textit{evaluation role} should use to generate feedback.
    
    \item \textbf{When:} Timing of AI Feedback. 
    We explored preferences for when \textit{ExReflect}’s feedback should be introduced, either interwoven with the \textit{experience role}’s enactment or delivered after it concludes, clarifying how the workflow can balance immersion and critique without disrupting creative flow.  
    
    \item \textbf{How:} Style and Transparency in AI Feedback.  
    Participants identified expectations for the tone and explanation type of feedback, informing how \textit{ExReflect} should present the \textit{evaluation role}’s feedback to ensure it is both transparent and actionable.  
\end{itemize}

\subsubsection{Data Analysis and Results}\label{sec:five-dimension framework}


Each interview lasted approximately 30 to 40 minutes. We conducted a qualitative thematic analysis, organizing findings into four design goals (DGs) that guide how \textit{ExReflect} supports reflection in screenplay refinement.

\textbf{DG1: High-Quality Experiential Simulation and Understandable Critical Feedback (Who, What, How).}
Participants distinguished between expectations for AI simulation and feedback. For simulation, they emphasized that the \textit{experience role} should function like a character, explicitly presenting inner thoughts (i.e., the personal experience generated by the \textit{experience role} in the given scenario) as the basis for reflecting the behavioral logic. For feedback, they expected the \textit{evaluation role} to act like an actor, stepping outside the enactment to critically examine the scene and articulate issues in the draft. Participants further highlighted the importance of pairing these immersive and credible inner thoughts with feedback from the \textit{evaluation role} that is specific and transparent. P1 noted, \textit{``Simultaneously displaying inner thoughts would be helpful''}, and P4 added, \textit{``This is the underlying character logic''}, while P3 stressed, \textit{``A reasoning path should be shown for each question.''}

\textbf{DG2: Alignment Between Simulation, Reflection, and Screenplay (What).}
\revise{Participants emphasized that feedback produced in the \textit{evaluation role} should closely match the way they reflected on their screenplays during refinement. Based on their accounts, we summarized this requirement as five dimensions that guide screenwriters' reflection in refinement:} character emotions, behavioral motivation, character relationships, plot pacing, and thematic consistency. They also stressed that such reflections should draw on contextual information rather than operate on isolated lines. For instance, P3 highlighted the need to \textit{``upload all existing information''} and \textit{``connect to the character’s backstory and personality''}. P7 described checking motivations by \textit{``reasoning backward from the story climax''}, while P8 noted the importance of \textit{``cross-referencing behaviors with character biographies''}. \revise{Together, these views indicate that screenwriters expect AI feedback to align with both the five dimensions and the information contained in their draft materials, ensuring that the feedback remains aligned with the narrative context they intend to express.}

\textbf{DG3: Stimulating Creative Insight and Refinement Willingness (What, How).}
\revise{We found that participants hope AI can surface moments where emotion, motivation, relationships, pacing, or theme become unclear or misaligned, giving them concrete entry points for reflection. They emphasized that such insights should critically expose blind spots rather than simply provide solutions or correctness judgments. For instance, P3 explained that when AI uncovers an inner meaning she had not noticed, it prompts her to ask \textit{``whether this meaning is what I intend''} and to assess whether refinement is needed. P5 noted that AI can reveal limits in her initial conception when writing unfamiliar topics, helping her adjust characters during refinement. P4 expected analyses of motivation to reveal inconsistencies that habitual writing patterns might overlook, prompting checks such as \textit{``whether this action is justified''} or \textit{``whether this line supports the character’s action and storyline''}. Taken together, these accounts show that screenwriters expect AI to highlight issues and offer starting points for reflection by providing insight and motivation for refinement.}

\textbf{DG4: Balancing Intervention Frequency and Creative Flow in Feedback Timing (When).}
Participants emphasized the importance of timing in feedback, calling for a balance between in-action and on-action approaches. Urgent, high-impact issues should be raised during enactment as \textit{instant feedback} to prevent cascading effects across the screenplay, while equally important but less urgent matters can be addressed afterward as \textit{post-hoc feedback} to minimize disruption to the creative flow.


\subsection{System Design}
Building on \textit{ExReflect} and the four DGs, we design \textit{DuoDrama} and detail how its backend (Secs.~\ref{sec:pre-processing} and~\ref{sec:dual_memory}) and frontend (Sec.~\ref{sec:frontend}) components are organized to support reflection in screenplay refinement.
\revise{Because screenplay refinement requires understanding how multiple characters jointly shape a scene, \textit{DuoDrama} uses the \textit{ExReflect} for each agent within a coordinated multi-agent architecture. After users upload screenplay excerpts through the frontend, the backend parses the file into scenes, identifies characters, and instantiates an \textit{ExReflect}-powered agent for each character, each holding its own persona profile, memory module, and reasoning pipeline. Each agent individually plays one character in the screenplay as the \textit{experience role} to generate inner thoughts as personal experiences, and then shifts to the actor portraying that character as the \textit{evaluation role} to generate feedback questions (feedback)\footnote{For simplicity, we use the term ``feedback'' to refer to ``generated feedback questions'' throughout the paper.}. The feedback will finally be displayed on the interface to support screenwriters' reflection.}

\subsubsection{Screenplay Pre-processing and Multi-Agent Orchestration}\label{sec:pre-processing}
To make the original screenplay uploaded by users usable for experience generation in the \textit{experience role}, it should first be structurally processed and converted into a unified format. The backend automates this conversion through a two-stage preprocessing pipeline, which is triggered by the upload panel. First, using a structured LLM prompt, the entire screenplay is divided into a list of discrete scenes, with scene boundaries identified based on the common scene heading format (such as ``INT. Coffee Shop - DAY''). Considering that some screenplay content may be filtered, a fallback mechanism based on regular expressions is adopted to identify common scene headings at the beginning of lines for scene segmentation. After the screenplay is divided into individual scene blocks, the second stage of preprocessing begins: we use a one-shot LLM Chain to guide the model in distinguishing character names and precisely separating character-related dialogues from action sequences, enabling subsequent line-by-line analysis. The original screenplay’s temporal and structural integrity is preserved to ensure that narrative elements align with subsequent agent responses.

\revise{After parsing the screenplay, we implement the system within a multi-agent architecture because screenplays typically involve multiple characters, assigning one agent to each character. A central orchestrator then processes the script line by line and activates the corresponding agent whenever that character speaks or acts. The active agent generates inner thoughts as personal experience based only on the public actions and dialogue available up to that point in the screenplay, and this experience is visible only to that agent. When control shifts to another agent, the system updates the public interaction context with the latest actions and dialogue in the screenplay. Through this coordination, each agent relies on the evolving public interaction context together with its own personal experience, rather than reasoning in isolation. Using personal experience as an additional interaction context when generating feedback is the mechanism we refer to as experiential grounding in this paper. This mechanism enables each agent to produce feedback independently based on its own generated experiences, while remaining contextual within the multi-agent architecture.}

\begin{figure*}
 \centering         
\includegraphics[width=\textwidth]{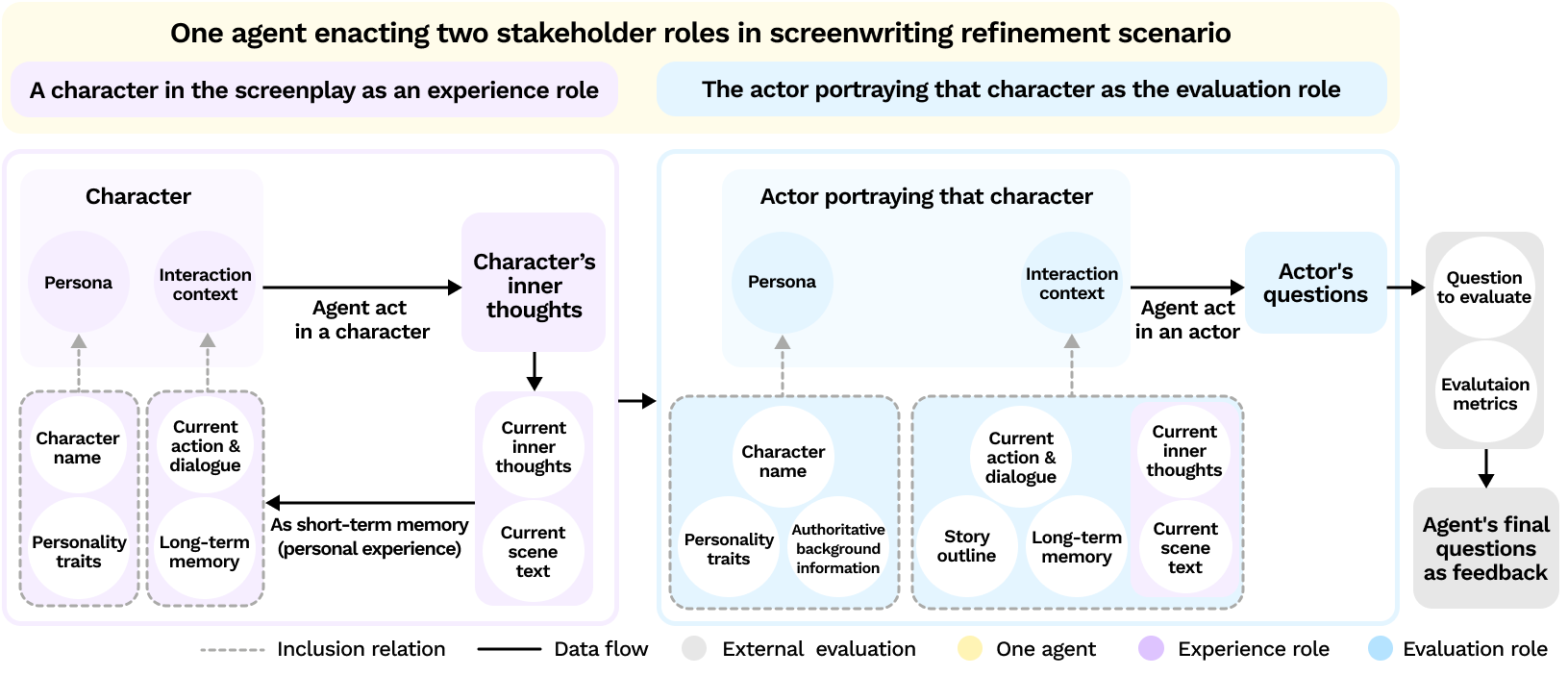} 
\caption{The detailed requirements for \textit{ExReflect} in the \textit{DuoDrama} system. A single agent enacts two complementary stakeholder roles within the screenplay refinement process. First, it acts as a character in the screenplay to generate \revise{the character’s current inner thoughts and its current scene text, where the current inner thoughts refer to this character produced in the present turn, and the current scene text aggregates all inner thoughts generated by this character in previous turns. Together, these form the agent’s short-term memory as personal experience (\textit{experience role}). Second, the agent shifts to the actor portraying that character to generate questions, using the personal experience as an additional input that supplements the interaction context (\textit{evaluation role}). The agent then applies the evaluation metrics to assess these questions, and those that satisfy the criteria become the final feedback presented to the user. \revise{For detailed evaluation metrics, please refer to Sec.~\ref{sec:dual_memory} and the supplementary material.}}
}
 \label{system_narrow}
 \end{figure*}

\subsubsection{ExReflect-powered Agent}\label{sec:dual_memory}

\revise{To implement each agent, \textit{DuoDrama} employs \textit{ExReflect}, which sequentially assigns two role perspectives to a single agent. The agent first adopts an \textit{experience role}, embodying a character in the screenplay to produce inner thoughts as personal experience. It then shifts to an \textit{evaluation role}, the actor portraying that character, to provide feedback based on these experiences. The system provides two complementary forms of feedback: \textit{instant feedback} at the sentence level and \textit{post-hoc feedback} at the scene level (see Fig.~\ref{system_narrow}).}

\textbf{Character Agent Initialization (DG1, DG2).}  
To support the role selection interface, we extract explicit character bios provided by the author from the screenplay, treating them as the factual foundation of each character. Since screenwriters often vary in how they write these bios, in terms of style and level of detail, we additionally implement a function that derives structured persona information, including background, core personality traits, goals, and motivations, by prompting the LLM to synthesize from both the scenes where the character appears and any authoritative bios if available. Each character agent is equipped with a dual-memory module comprising long-term and short-term memory. Long-term memory stores past event descriptions from previously enacted scenes that the agent previously experienced, which are encoded into high-dimensional vectors for later retrieval. During simulation, when an agent recalls relevant experiences, the current situation (i.e., the current action and dialogue) is embedded as a query vector, and the retriever conducts a similarity search to retrieve semantically related memory vectors. Short-term memory, by contrast, functions as a dynamically maintained context window that stores the agent's personal experience, specifically aggregating the current scene text and current inner thoughts,  preserves the most recent events within a scenario as real-time references for enactment. This mechanism integrates the character’s subjective perspective: each agent can access its own internal state (including previous inner thoughts) while perceiving only the publicly observable actions and dialogues of other characters. By preserving this information boundary, the system ensures that agents respond realistically to visible events during scene progression while maintaining their independent perspectives.


\textbf{\revise{Experience-Grounded Feedback Generation} (DG1, DG2, DG3, DG4).}  
This algorithm operationalizes the \textit{ExReflect} workflow through a two-stage prompt architecture that links enactment with feedback via stakeholder role switching. In the first stage, the agent enacts \textit{experience role}, embodying a specific character in the screenplay, to generate personal experience. The agent perceives the available information and reasons step by step through a Chain-of-Thought process: (i) initial interpretation, formulating the character's intuitive reaction to recent events and dialogues; (ii) memory recall, linking this reaction with the character’s background and traits to analyze why such feelings emerge; (iii) objective definition, articulating the immediate but unspoken goal in the interaction; and (iv) synthesis, condensing the reasoning into concise first-person inner thoughts that reveal the emotions and rationale driving the character's actions or dialogue. This enactment yields situated personal experience grounded in the perspective of the screenplay character. In the second stage, the agent shifts to \textit{evaluation role}, taking the position of the actor who performed that character before, to generate feedback based on the personal experience generated in \textit{experience role} as additional short-term memory. The \textit{evaluation role} applies five dimensions, covering emotions, behavioral motivations, character relationships, plot rhythm, and thematic consistency, to identify potential issues and retrieve supporting textual evidence, generating feedback. 
\revise{Besides, to ensure the feedback aligns with an actor's natural spoken expression, we incorporate language style rules: the agent is instructed to maintain a fluent, colloquial tone and to avoid subject-pronoun confusion. To further foster expression diversity, it restricts formulaic sentence structures (e.g., ``While performing...'' or  ``When I played...''), academic jargon, and over-reliance on specific phrases (e.g., ``could it be''). Furthermore, the agent refrains from simply copying original scene text or settings to avoid repeating known information.}

\revise{\textbf{\revise{Experience-Grounded Feedback }Evaluation (DG1, DG2, DG3, DG4).}}
\revise{To ensure that the system raises the right \revise{feedback} at the right time, we designed a self-verification mechanism implemented through a dedicated evaluation prompt. This mechanism operates outside the roles in the agent, serving as a neutral check that assesses and filters the system’s generated feedback before it is presented to the user.}
\revise{Specifically, to evaluate \textit{instant \revise{feedback}}, we instruct the model to assess each candidate \revise{feedback} according to four criteria.
(a) \textit{Evidence Verification:} To mitigate hallucinations, an issue frequently noted in co-creative AI \cite{10.1145/3173574.3174223}, the rationale must be grounded in unambiguous evidence from the user-provided authoritative background, story outline, relevant memories, current scene text (personal experience), or current action or dialogue. Any mismatch or unverifiable claim results in rejection.
(b) \textit{Expression Diversity:} 
Prior work shows that writers aim to avoid clichés and value support that introduces fresh semantic associations to counter creative fixation~\cite{10.1145/3290605.3300526}. To support this, we embed expression-diversity rules directly into the feedback-generation prompt and then apply a secondary check during evaluation. The mechanism uses the model’s semantic reasoning to balance stylistic variety with feedback quality, detecting issues such as jargon stacking or templated openings as described in the generation prompt.
This criterion functions as a soft penalty that interacts with the other three criteria. When all four criteria, including expression diversity, are met, the system delivers the feedback. When only expression diversity is violated, the evaluator trades off expression deviation against usefulness: feedback that provides clear insight may still pass, whereas low-value feedback is filtered out. If any of the other criteria are not satisfied, the feedback is rejected regardless of expression diversity, and when both expression diversity and at least one other criterion fail, the feedback is discarded.}
\revise{
(c) \textit{Five Target Dimensions:} 
The \revise{feedback} must explicitly address at least one dimension among the five dimensions \revise{(refer to \ref{sec:five-dimension framework})}, including character emotions, behavioral motivation, character relationships, plot pacing, or thematic consistency, to ensure relevance to the screenplay reasoning framework.
(d) \textit{Impact and Timing:} 
The evaluation estimates whether omitting the \revise{feedback} would cause a motivational break, a logic gap, or weaken the upcoming narrative. Guided by findings that feedback timing influences user perception \cite{10.1145/3635636.3656183,10.1145/985692.985727}, \textit{instant feedback} is only accepted when the issue is judged to have a severe impact requiring immediate resolution.
\revise{Feedback} with minor or deferrable effects are withheld to avoid disrupting creative momentum.}
\revise{
To evaluate \textit{post-hoc \revise{feedback}}, the same four core criteria are applied with considering the full screenplay and enriched personal experience. The rationale must again be supported by reliable evidence, but the focus shifts to issues of severe impact that do not require immediate interruption.
Instead of focusing on immediate interruptions, the evaluation prioritizes feedback that addresses the scene at a holistic level. Feedback focused on specific actions or dialogue is rejected unless it influence the entire scene.
After applying these constraints, the evaluation chain issues a binary decision: only feedback that satisfies all criteria is retained and shown to the user. 
A complete version of the evaluation prompt and examples comparing rejected and adopted \revise{feedback} are provided in the supplementary material.}



\subsubsection{User Interface Design}\label{sec:frontend}
\begin{figure*}
    \centering
    \includegraphics[width=\textwidth]{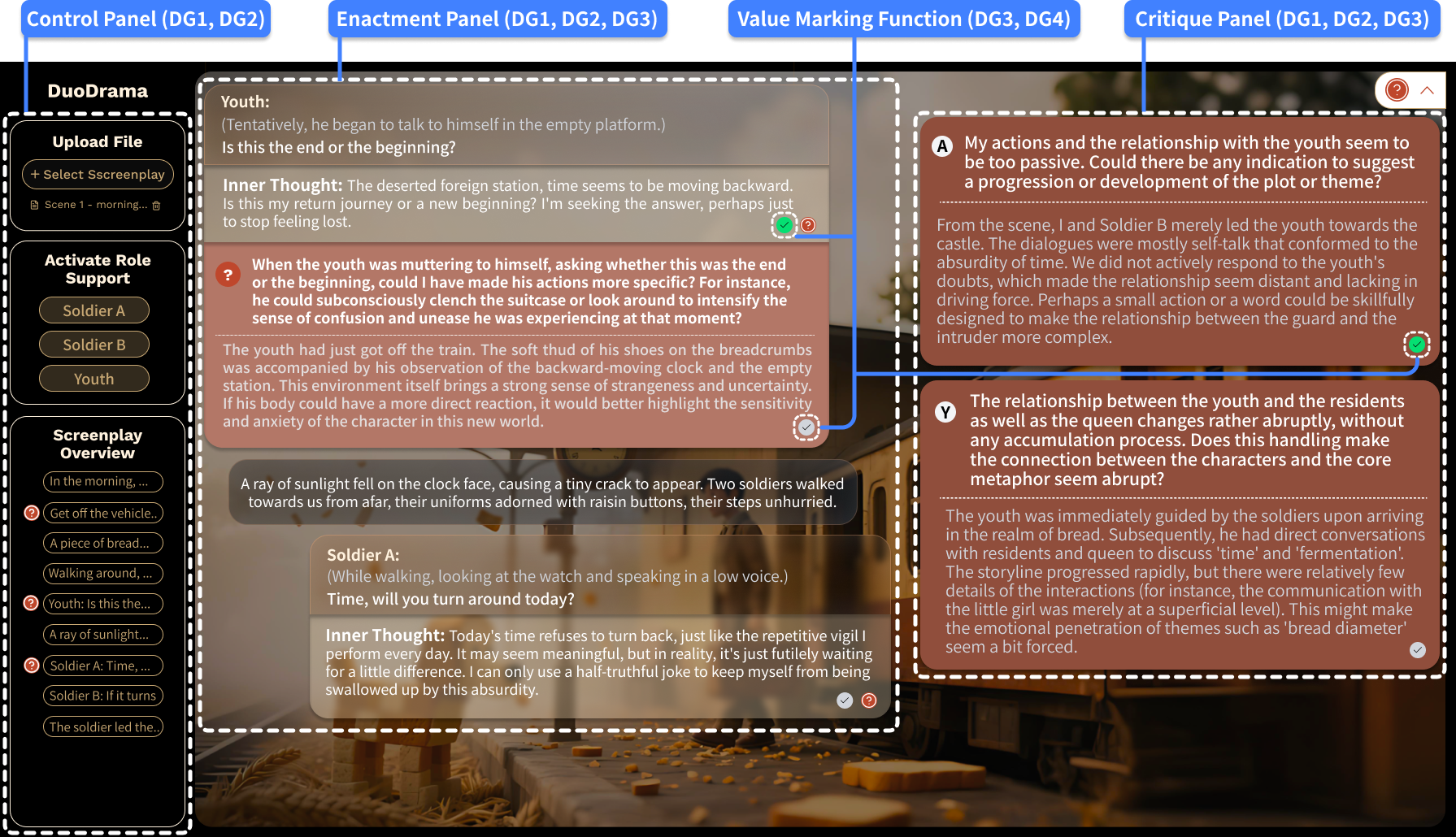}
    \caption{\revise{The panels in the frontend interface of the \textit{DuoDrama}, mapped to design goals (DGs). \textbf{Control Panel} manages screenplay materials and roles to set the stage for high-quality simulation (DG1) and ensure screenplay alignment (DG2). \textbf{Enactment Panel} (DG1, DG2, DG3) presents line-by-line dialogues and inner thoughts to support high-quality simulation (DG1) and contextual alignment (DG2), while triggering creative insight and refinement willingness (DG3) via \textit{instant feedback}. \textbf{Critique Panel} (DG1, DG2, DG3) lists \textit{post-hoc feedback} to facilitate critical understanding (DG1) and checks contextual alignment (DG2) to trigger creative insight and motivate refinement willingness (DG3). \textbf{Value Marking Function} (DG3, DG4) enables users to identify and retain useful AI outputs for refinement (DG3), while being designed to balance intervention timing with creative flow (DG4).}}
    
    \label{fig:frontend_interface}
\end{figure*}

Based on these algorithms, the frontend interface contains four panels (see Fig~\ref{fig:frontend_interface}).

\textbf{Control Panel (DG1, DG2).}  
The control panel serves as the entry point for managing materials and role enactments. It provides three main functions:
(i) \textit{Uploading screenplay files} through the \textit{+ Select Screenplay} button. Uploaded materials (e.g., scenes, character biographies) are automatically parsed into structured segments for subsequent enactment and critique.  
(ii) \textit{Assigning roles} via the \textit{Activate Role Support} buttons. By clicking on a character (e.g., Soldier A, Soldier B, Youth), the system instantiates this role to produce inner thoughts and experience-grounded \revise{feedback}. Users can flexibly select different combinations of roles for enactment.  
(iii) \textit{Navigating screenplay lines} with the \textit{Screenplay Overview} function. Each utterance or action in the uploaded scene is displayed as a clickable index. Clicking on any line scrolls the enactment panel to the corresponding position, ensuring efficient access and contextual continuity.  
Together, these functions clarify the narrative scope, preserve completeness, and support precise navigation.  

\textbf{Enactment Panel (DG1, DG2, DG3).}  
The enactment panel simulates a natural reading rhythm by presenting dialogues and inner thoughts line by line. Roles selected in the control panel are enacted with generated personal experience (i.e., action or dialogue with inner thoughts), while unselected roles are shown directly from the screenplay (i.e., action or dialogue). \textit{Instant feedback} appear as red icons placed beside specific sentences. When users click on the icon, it expands into a collapsible box that presents targeted, \textit{instant} \revise{feedback} nested beneath the corresponding line. Users can close the box by clicking again, minimizing disruption to the narrative flow. This design makes \revise{feedback} available at precise textual locations.

\textbf{Critique Panel (DG1, DG2, DG3).}  
The critique panel presents \textit{post-hoc \revise{feedback}} in a bubble-style list, highlighting issues that extend beyond individual lines. Users can scroll through the panel to examine these broader prompts. By separating \textit{instant \revise{feedback}} in the enactment panel from \textit{post-hoc \revise{feedback}} in the critique panel, the system enables both fine-grained inspection and higher-level narrative reasoning without introducing visual clutter.


\textbf{Value Marking Function (DG3, DG4).}  
The value-marking function allows users to identify and retain useful AI-generated content. By clicking the green checkmark icon attached to inner thoughts, \textit{instant feedback}, or \textit{post-hoc feedback}, users can mark selected items as valuable, which are then visually highlighted for easy recognition across panels. When an item is marked, the system aggregates it with its contextual metadata, including the associated character, scene content, scene number, and feedback type, into a structured JSON object that is stored in the background. This mechanism enables users to efficiently revisit and locate relevant information and supports the capture of revision-relevant insights while filtering out less useful content during later review.


\subsection{System Walk-Through}

\begin{figure*}
    \centering
    \includegraphics[width=\textwidth]{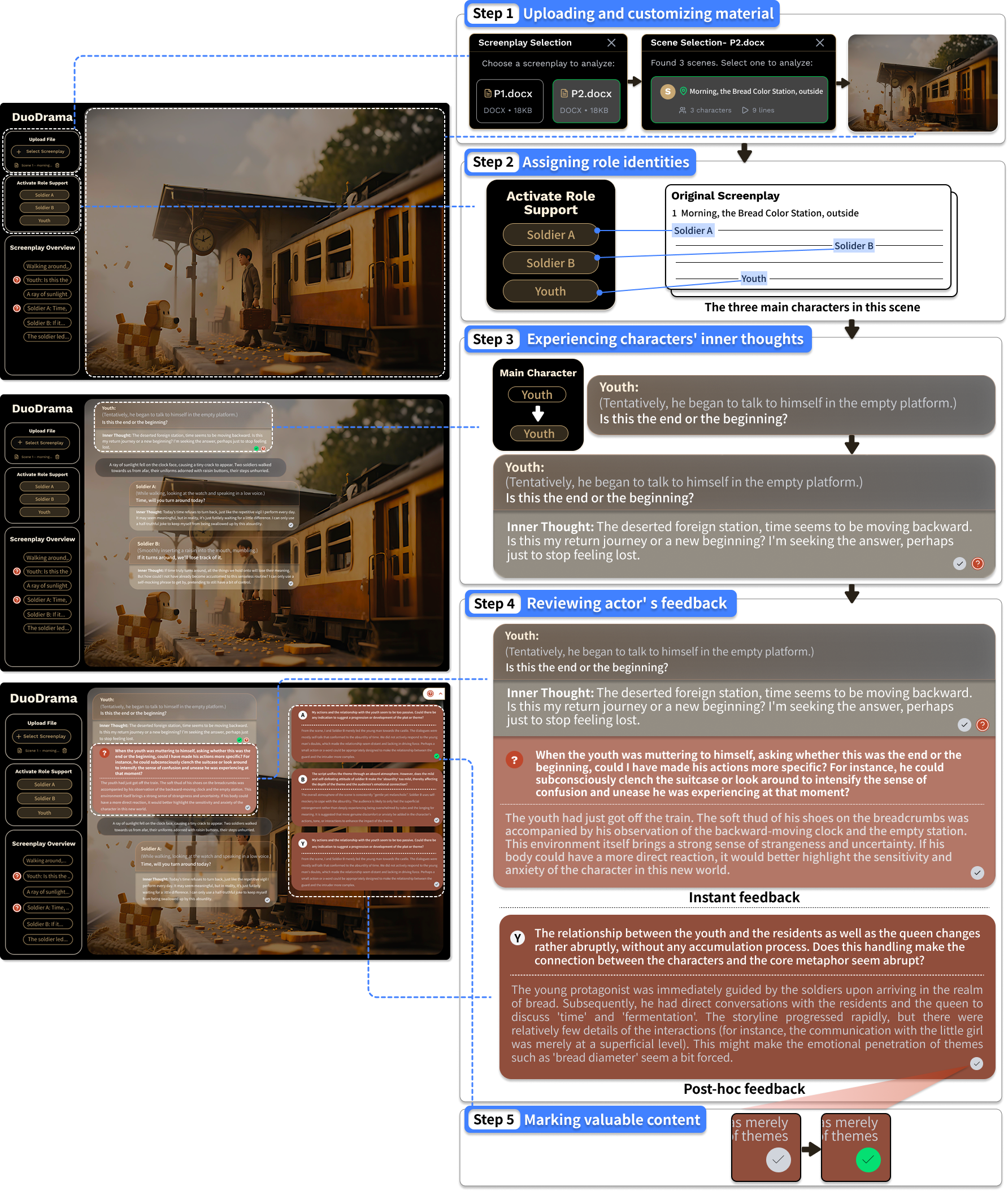}
    \caption{The Process and Features of the \textit{DuoDrama}. Step 1: Uploading and customizing material. Step 2: Assigning role identities. Step 3: Experiencing characters' inner thoughts. Step 4: Reviewing actor' s feedback \revise{feedback}. This includes: (a) \textit{instant \revise{feedback}} that appear directly below relevant inner thoughts, and (b) a series of \textit{post-hoc \revise{feedback}} provided after the scene simulation. Step 5: Marking valuable content.}
    \label{fig:system_process}
\end{figure*}

\revise{To illustrate how \textit{DuoDrama} supports screenwriters' reflection, we describe the experience of an imagined user, Daisy. This narrative is grounded in the interaction shown in Fig.~\ref{fig:system_process}, which is drawn from participant P12’s use of the system.}

\textbf{Step 1: Uploading and customizing material.}
\revise{Daisy begins by uploading her screenplay file. The system automatically parses the document into scene lists. As shown in Fig.~\ref{fig:system_process}, three scenes appear, each with a thumbnail and a short description. Daisy chooses the scene ``Morning, the Bread Color Station, outside’’ as her target. She then adds an image of the platform with a dog and a man. This visual reference appears on the right panel and helps her keep track of the atmosphere of the scene throughout later steps.}

\textbf{Step 2: Assigning role identities.}
\revise{After selecting the scene, Daisy reviews the characters detected in this part of the screenplay. The system identifies Soldier A, Soldier B, and the Youth. Daisy confirms the list and moves to the enactment stage.}

\textbf{Step 3: Experiencing characters’ inner thoughts.}
\revise{Daisy chooses whose inner thoughts she would like to see during enactment. She selects the Youth because the scene focuses on his emotional state, and she also selects Soldier A and Soldier B to understand how they interpret the same moment. The system displays the screenplay line by line at the bottom left for an overview. The enactment panel then shows the original screenplay text together with the selected characters’ inner thoughts. In Fig.~\ref{fig:system_process}, the Youth mentions whether he is standing at an end or a beginning and describes the sense of time and space slipping backward. Daisy reflects and compares this with her intent. She reflects that she had meant to signal a quiet form of fear, but the Youth’s internal voice expresses confusion rather than fear. This prompts her to reflect on whether her draft lacks the cues needed to guide readers toward the emotion she wants. She notes to herself that a small revision in the Youth’s physical reaction might better anchor the desired tension.}

\textbf{Step 4: Reviewing actor’s \revise{feedback}.}
\revise{As Daisy continues reading the inner thoughts, small red icons appear next to several lines. Clicking an icon reveals an \textit{instant \revise{feedback}} tied to that moment. One \revise{feedback} asks whether the Youth’s hesitation could be shown through a clearer physical action, such as a pause before stepping forward or a repeated glance toward the station sign. This supports Daisy’s reflection: she reflects that the script currently asks the audience to infer too much from sparse dialogue, and that adding one simple action may help communicate his internal struggle. Another \textit{instant feedback} asks whether the dog in the visual reference might be used as a grounding point for the Youth’s shifting emotions. Daisy reflects that she had treated the dog only as background, but this suggestion helps her consider using it as a stabilising element to contrast with the Youth’s uncertainty. Daisy then opens the critique panel to view broader \textit{post-hoc \revise{feedback}}. One \revise{feedback} highlights that the relationship between the Youth and the residents changes suddenly in the current draft, which may weaken the emotional progression of the scene. This encourages Daisy to reflect not only on single lines but on how her scene moves from isolation to brief connection. She reflects that she needs one transitional beat to prepare for the shift. Through these instant and post-hoc \revise{feedback}, Daisy receives support for both local and scene-level decisions, triggering reflection for her screenplay refinement.}

\textbf{Step 5: Marking valuable content.}
\revise{Finally, Daisy marks content she wants to retain by clicking the green checkmark next to particular inner thoughts or \revise{feedback}. In this example, she keeps feedback related to refining the Youth’s physical behaviour, using the dog as an emotional anchor, and smoothing the transition in the Youth’s relationship with the residents across scenes. Marked items remain highlighted, allowing her to revisit them easily. These marked points represent concrete evidence of her reflection and guide her potential refinement.}

\subsection{\textit{DuoDrama} System Implementation} \label{sec:Implementation}

\revise{We employ \textit{ExReflect} as a single-agent workflow that shifts from an \textit{experience role} to an \textit{evaluation role} in \textit{DuoDrama}. The agent first performs enactment to generate inner thoughts as personal experience, which are stored as short-term memory. These are then used as additional context when the agent shifts to an external, actor-like perspective to generate feedback. To operationalize this mechanism, we implement a dual-memory module and a structured LCEL pipeline\footnote{LangChain Expression Language (LCEL): \url{https://python.langchain.com/docs/expression_language/}}. During enactment, the LCEL inner-thoughts chain takes the current situation and persona profile as input and outputs JSON-structured thoughts, which are written into short-term memory. Before feedback generation, the system retrieves relevant long-term traces through FAISS\footnote{Facebook AI Similarity Search (FAISS): \url{https://github.com/facebookresearch/faiss}} and merges them with short-term memory to form a synthesized context window. The feedback generator chain then produces JSON-structured feedback through a Chain-of-Thought process, ensuring that feedback remains grounded in enacted experience.}

\revise{Because screenplay refinement involves multiple characters shaping a scene, \textit{DuoDrama} integrates several ExReflect-powered agents within a coordinated multi-agent architecture. For each screenplay character, the system instantiates an agent that runs its own LCEL enactment pipeline, dual-memory module, and feedback generator. Each agent alternates between a character's perspective for enactment and an actor's perspective for evaluation. This architecture enables character-specific experiential grounding while maintaining consistent coordination across multiple agents.}

\revise{\textit{DuoDrama} is implemented as a web application with a FastAPI backend exposing REST endpoints and a Next.js/React frontend for interaction. The backend orchestrates all agent behaviors through LCEL pipelines, manages memory modules, and performs FAISS-based retrieval. The frontend displays each character's inner thoughts, retrieved memory traces, and the reasoning supporting each feedback. All LLM reasoning uses GPT-4.1 via the Azure OpenAI API, and GPT-4o is used to produce paired visualizations that accompany enactment outputs.}

\section{User Study}  

To evaluate whether our system achieves the four DGs, we conducted a two-session user study with fourteen professional screenwriters. The study employed a mixed-methods approach, combining qualitative and quantitative analyses to capture both user experience and comparative evaluations. The study protocol was reviewed and approved by our institution's ethics committee (IRB).  

\subsection{Study Design}
Our user study employed a two-session design to evaluate the \textit{DuoDrama} against the four DGs. Session~1 captured participants' experiences with \textit{DuoDrama}, while Session~2 provided comparisons across four conditions. Together, they offer complementary results in both the user experience and the advantages of the \textit{DuoDrama}.  


\subsubsection{Session 1: Experience and Effect Test}
Session~1 examined participants' experiences with \textit{\textit{DuoDrama}}, covering DG1, DG2, DG3, and DG4. Participants first interacted with the system on their chosen \textit{Scene A} from the screenplay they provided, encountering both AI-generated inner thoughts and feedback. They then completed a 13-item Likert-scale questionnaire (1–7) evaluating the quality of simulation (DG1), the alignment of feedback (DG2–DG3), and the appropriateness of feedback timing (DG4). 
The items are available in Fig.~\ref{fig:quantitative_results}. We also measured the 10-item \textit{System Usability Scale} (SUS, 1–5). Following the questionnaire, participants took part in a semi-structured interview that probed their reasoning behind ratings, preferences for \textit{instant feedback} versus \textit{post-hoc feedback}, and perceptions of whether feedback timing supported or disrupted their creative flow. 

\subsubsection{Session 2: Offline Comparative Evaluation}  
Session~2 served as the primary evaluation, designed to address DG1, DG2, and DG3. Participants were presented with four systematically varied AI feedback conditions, shown in randomized order. This session was based on a new excerpt (\textit{Scene B}), which was another scene they had selected from their own screenplay for refinement.

\begin{enumerate}
    \item \textbf{\revise{Eval-PE (\textit{DuoDrama})}: \textit{Evaluation role} perspective with \textit{experience role}'s personal experience (PE).} The agent first simulated the \textit{experience role}, enacting a character to generate inner thoughts as personal experience (PE), and then shifted to the \textit{evaluation role}, providing suggestions from the perspective of the actor portraying that character. The five dimensions (refer to Sec \ref{sec:five-dimension framework}) were applied to both \textit{instant feedback} and \textit{post-hoc feedback}, with its rationales grounded in the PE as additional context.
    \item \textbf{\condtwo{}: \textit{Experience role} perspective with PE.} The agent simulated the character as the \textit{experience role} to generate PE and posed first-person feedback directly from the character’s perspective, with its rationales grounded in the PE as additional context. Similar to \condone, the five dimensions guided both \textit{instant feedback} and \textit{post-hoc feedback}. The difference is that the agent was restricted to information knowable to the character and could not access the full screenplay beyond that \textit{experience role}'s perspective.  
    \item \textbf{\condthree{}: \textit{Evaluation role} perspective without PE.} The agent generated feedback from an \textit{evaluation role} perspective without performing any simulation. Since no \textit{experience role} in this condition, no inner thoughts as PE was generated, there were no line-specific cues to anchor \textit{instant feedback}. As a result, the feedback only surfaced as \textit{post-hoc feedback} derived from the five dimensions.  
    \item \textbf{\condfour{}: Screenplay reviewer perspective without PE.} To reflect industry-standard screenplay review practices, the agent relied solely on the screenplay content and generated feedback from an external reviewer's perspective. Because no \textit{experience role} was present in this condition, no PE was involved. As a result, the system could not generate line-anchored \textit{instant feedback} and instead delivered only \textit{post-hoc feedback} along the five dimensions. This condition served as a screenwriting industry baseline.
\end{enumerate}  

\revise{These four conditions were designed to disentangle the separate and combined effects of personal-experience (PE) grounding and feedback perspective, allowing us to evaluate which mechanisms in \textit{DuoDrama} meaningfully contribute to the screenwriters' reflection in refinement. First, \textbf{\condone vs.~\condtwo} isolates the effect of feedback perspective by comparing the \textit{evaluation role} and \textit{experience role} when both are grounded in PE. Next, \textbf{\condone vs.~\condthree} examines whether PE grounding itself strengthens feedback generation within the same \textit{evaluation role} perspective, thereby evaluating the core principle of experiential grounding in \textit{DuoDrama}. Finally, \textbf{\condone vs.~\condfour} compares \textit{DuoDrama}'s full mechanism against the industry baseline of professional screenplay reviewing, testing whether the system provides advantages over standard feedback practices.}

\revise{After reviewing each condition, participants rated it on 18 items using 7-point Likert scales. These items measured alignment (DG2), quality (DG1), perceived effectiveness (DG3), and user reflection (DG3). Quality (DG1) assesses whether the feedback offers sufficient content richness, clarity, relevance, comprehensibility, and specificity of detail. High-quality feedback provides concrete cues that help screenwriters understand why a moment may not work as intended, which in turn facilitates reflection. Alignment (DG2) evaluates whether the feedback remains consistent with the narrative context across character emotion, behavioral motivation, character relationship, plot pacing, and thematic consistency. When feedback stays aligned with the story context, screenwriters can trust that the feedback meaningfully relates to their creative intentions, which reduces the effort required to interpret or reconcile the feedback and supports reflection. Perceived Effectiveness (DG3) subdimensions, including emotional insight, motivational insight, relationship insight, plot pacing insight, thematic insight, and revision motivation, capture whether the feedback helps screenwriters notice overlooked issues and encourages them to reconsider their decisions. These dimensions correspond to how screenwriters structure their own reflection during refinement. Therefore, the quantitative scores provide direct evidence for how effectively each condition supports human reflection. In addition, we incorporated the depth and richness subdimensions from user reflection (DG3) to further assess how the system enhances reflection. These measures strengthen the validation of the system’s ability to deepen and broaden users’ reflection~\cite{mezirow1990fostering}. Follow-up interviews then contextualized the quantitative results by allowing participants to explain their preferences, identify high-quality examples, and articulate perceived differences across conditions. Together, the quantitative and qualitative analyses show how each design mechanism contributes to \textit{DuoDrama}.}

\subsection{Study Participants and Procedure}
We recruited fourteen professional screenwriters, each with prior experience using AI tools in screenwriting practice (see Fig~\ref{tab:participants_profile}). This background allowed participants to make informed comparisons between our system and existing AI supports. \revise{All participants submitted a complete screenplay draft, including an outline, character descriptions, and the full screenplay with actions and dialogues, and they were preparing these drafts for further refinement. Screenplays with any missing sections were not accepted, which ensured that all participants provided the same types of information, even though the stories themselves differed. As shown in the Sec.~\ref{sec:dual_memory} and backend algorithm figure in the supplemental material, we then applied an automated screenplay pre-processing step to standardize the content and format before system processing.} The user study consisted of two sessions. In session 1, participants interacted with our system using their screenplay excerpt (\textit{Scene A}) to experience the \textit{DuoDrama} workflow. In session 2, they completed an offline comparative evaluation task with a different excerpt (\textit{Scene B}) to assess four conditions. Both sessions were conducted consecutively via an online meeting platform and lasted approximately 110 to 120 minutes in total for each participant.

\begin{table*}[ht!]
\centering
\caption{Demographic and professional information of the participants.
\textbf{Frequency} of AI use is rated on a 5-point scale: 1 (Unused), 2 (Occasionally use, $\geq$ once per month), 3 (Sometimes, 1-3 times per month), 4 (Often, at least once per week), and 5 (Frequent daily use, in almost every writing session).
\textbf{Screenwriting Experience} refers to the participant's total years of experience in screenwriting. \textbf{AI Usage} indicates the total duration of experience using AI in the screenwriting process. }
\label{tab:participants_profile}
\footnotesize
\renewcommand{\arraystretch}{1.3}
\begin{tabularx}{\linewidth}{@{} l l c c c c X @{}}
\toprule
\textbf{ID} & \textbf{Gender} & \textbf{Age} & \textbf{Screenwriting Experience (Years)} & \textbf{AI Usage (Years)} & \textbf{Frequency (1-5)} & \textbf{AI Platforms (e.g., ChatGPT)} \\
\midrule
P1  & Female & 25 & 8   & 1 & 3 & ChatGPT \\
P2  & Female & 32 & 12 & 1 & 4 & ChatGPT, DeepSeek \\
P3  & Female & 33 & 12  & 2 & 4 & ChatGPT, DeepSeek \\
P4  & Female & 19 & 2   & 2 & 3 & Deepseek \\
P5  & Male   & 29 & 9   & 1 & 5 & ChatGPT, DeepSeek \\
P6  & Female & 29 & 7   & 2 & 5 & ChatGPT \\
P7  & Female & 33 & 15  & 1 & 5 & Deepseek \\
P8  & Female & 39 & 15  & 1 & 4 & Deepseek \\
P9  & Female & 30 & 12  & 2 & 4 & ERNIE, Qwen, DeepSeek, Gemini, Doubao \\
P10 & Male   & 39 & 10  & 1 & 5 & ChatGPT \\
P11 & Male   & 33 & 10  & 2 & 5 & ChatGPT, ERNIE, Qwen, DeepSeek, NovelAI, Sudowrite \\
P12 & Female & 26 & 3   & 1 & 2 & ChatGPT \\
P13 & Male   & 27 & 10  & 1 & 2 & ChatGPT, ERNIE, DeepSeek \\
P14 & Female & 22 & 10  & 4 & 3 & ChatGPT, DeepSeek \\
\bottomrule
\end{tabularx}
\end{table*}


\subsection{Data Collection and Analysis}

Data collection combined quantitative scales and qualitative interviews across both sessions. Prior to the user study, a priori power analysis was conducted using G*Power (version 3.1.9.7) to determine the minimum sample size required for the study. The power analysis for a Wilcoxon signed-rank test (matched pairs) revealed that a minimum sample size of 13 was necessary to detect a large effect size ($d_z = 0.90$) with 80\% power at a significance level of $\alpha = 0.05$ (two-tailed). In our study, a total of 14 participants were recruited, which exceeded the minimum benchmark required by the power analysis. Therefore, the sample size of this study is sufficient to ensure the rationality and validity of the experimental results.

In Session~1, \textit{DuoDrama}'s usability was assessed using the System Usability Scale (SUS). Individual scores were calculated for each participant, and the average SUS score was then determined. A score of $\geq 68$ was considered indicative of good usability \cite{sauro2016quantifying}. User experience with the system's functionality was rated on a 7-point Likert scale. These data were summarized by calculating the proportional distribution of responses for each question and visualized using stacked bar charts. Semi-structured interviews were transcribed and analyzed thematically~\cite{braun2006using}, focusing on participants' perceptions of usability, user experience, and timing. In Session~2, ordinal ratings comparing different conditions were analyzed across multiple sub-dimensions. Pairwise comparisons between conditions (\condone vs.~\condtwo, \condone vs.~\condthree, and \condone vs.~\condfour) were conducted using the non-parametric Wilcoxon signed-rank test. Effect sizes were reported as $r$ (calculated as $Z/\sqrt{n}$), and two-sided $p$-values were used for significance testing. For each condition, the median score was reported. Qualitative interview data were used to supplement the quantitative findings, providing reasons why participants favored certain conditions.

\subsection{Study Session 1 Results: Experience and Effect Test}\label{sec:Session 1}

\begin{figure*}[t]
    \centering
    \includegraphics[width=\textwidth]{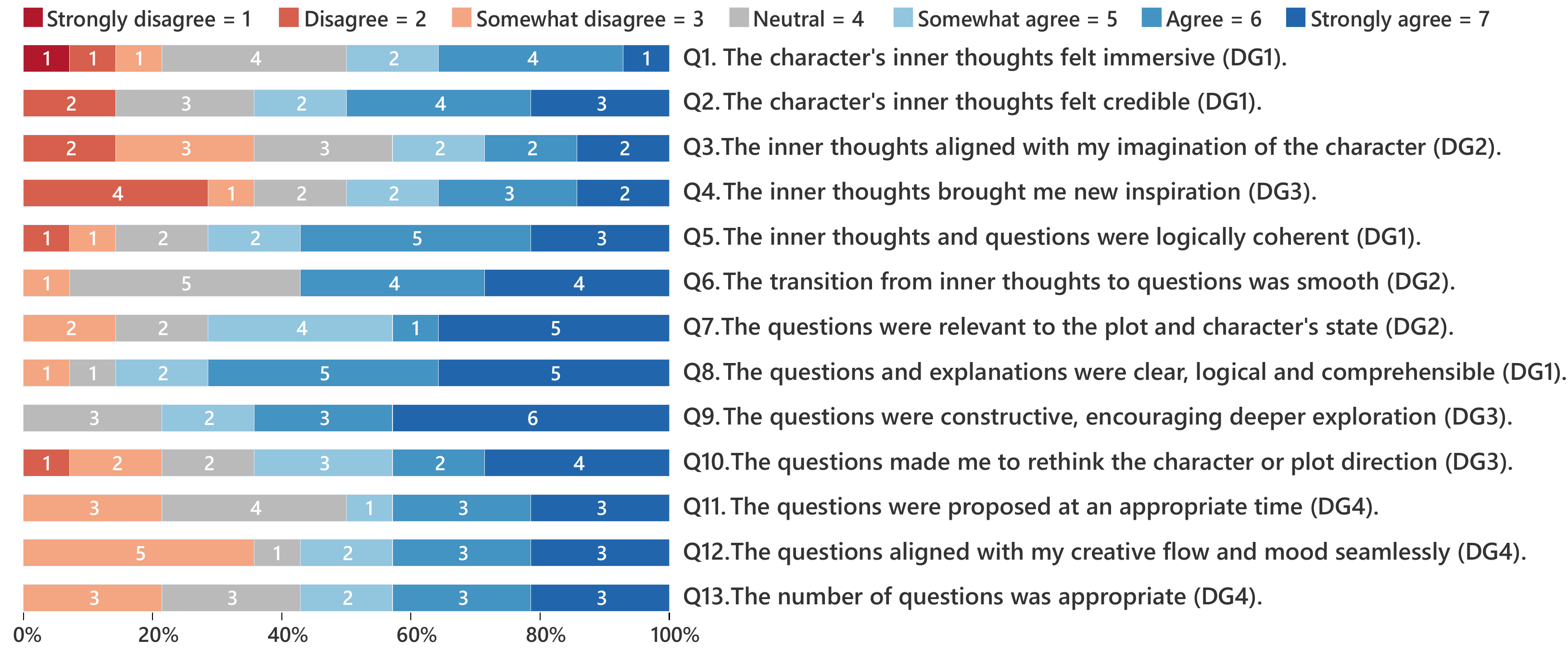}
    \Description{Results of the user experience survey in Session 1, showing participant ratings on a 7-point Likert scale, grouped by four DGs.}
    \caption{
    Results of the user experience survey in Session 1, showing participant ratings on a 7-point Likert scale. 
    The questions are grouped by four DGs: 
    (a) DG1 - High-Quality Embodied Simulation and Understandable Critical Feedback (Q1, Q2, Q5, Q8); 
    (b) DG2 - Alignment Between Simulation, Reflection, and Screenplay (Q3, Q6, Q7); 
    (c) DG3 - Stimulating Creative Insight and Refinement Willingness (Q4, Q9, Q10); and 
    (d) DG4 - Balancing Intervention Frequency and Creative Flow in Feedback Timing (Q11, Q12, Q13).
    \revise{In this figure, ``questions'' in the survey items refer to the feedback questions generated by \textit{DuoDrama}}.}
    \label{fig:quantitative_results}
\end{figure*}

The results from Session~1 addressed the four DGs and corroborated the general impressions formed by participants after using the system (see Fig.~\ref{fig:quantitative_results}). All participants agreed that \textit{DuoDrama} provided more professional and in-depth analysis than the AI systems they had previously used, and that the visible enactment of inner thoughts enhanced their trust in the subsequent feedback. Nine participants perceived the feedback as experience-grounded, making it more authentic and persuasive. Five participants further emphasized that this actor-like perspective filled a gap in their creative process by offering insights otherwise absent. Ten participants preferred \textit{post-hoc feedback} raised at the end of a scene, as these focused more on overall structure, character relationship, and theme, thereby prompting more directional reflection. In contrast, \textit{instant feedback} was acknowledged as valuable, but its excessive specificity and granularity sometimes led participants to doubt its practical usefulness for broader screenplay refinement. At the same time, thirteen participants emphasized the system's ability to surface overlooked details and provide new perspectives, which frequently triggered reflection for potential refinement. Overall, the system received an average SUS score of 84.46, well above the benchmark of 68~\cite{sauro2016quantifying}. These results indicate high usability along with broadly positive evaluations of the interaction experience. In the following sections, we detail how the system specifically addressed each DG through participants' evaluations, as shown in Fig.~\ref{fig:quantitative_results}.

\subsubsection{\revise{DuoDrama Provides High-Quality Logic and Clarity Feedback for Reflection}}
\revise{Results showed that the system performed well on credibility (Q2), logical coherence (Q5), and explanatory clarity (Q8), addressing the needs captured in DG1. For credibility, most participants gave positive scores\footnote{For simplicity, we use the term ``positive scores'' to refer to ratings of 5, 6, or 7.}, with 9 out of 14 selecting 5, 6, or 7. For logical coherence, 10 out of 14 participants selected positive scores. For explanatory clarity, 12 out of 14 participants selected positive scores. These results indicate that most participants viewed the feedback’s clarity, psychological reasoning, and structural consistency as strong. Users frequently praised the credibility, clarity, and logical quality of the system's feedback. Because clear and coherent reasoning helps screenwriters trace why a moment succeeds or fails, such feedback supports reflection in refinement by giving screenwriters concrete and interpretable cues to reconsider their decisions.} P14 remarked \textit{``The internal and emotional logic were coherent, so it is credible.''} P4 added \textit{``Credibility was partly because some overlapped with my expectations of inner thoughts, and another part was beyond my expectations, but still reasonable.''} P9 said \textit{``The biggest advantage is that it speaks human language. Unlike most AI nowadays, it does not give incomprehensible psychological analysis…like a person communicating with you...It captured the right points, the vocabulary was accurate.''} By contrast, ratings for immersion (Q1) showed more variability. \revise{For immersion, 7 out of 14 participants selected positive scores.} Several participants felt that although the enactments were coherent, they were less effective at conveying deeper emotional richness or layered psychological states, which reduced their sense of immersion. P13 commented that the AI was like \textit{``a good student at solving exam questions,''} with strong logical reasoning but less instinctive nuance. P2 described it as \textit{``a bit shallow,''} and P6, whose screenplay involved intricate psychological states, noted \textit{``for this kind of situation requiring complex psychology, the AI did not capture them fully.''} These findings suggest that while the system generally maintained credibility, coherence, and clarity, its ability to handle immersion with more complex or layered psychological content varied across participants' perceptions.

\subsubsection{\revise{DuoDrama Supports Smooth Interaction and Contextually Aligned Feedback for Reflection}}
\revise{In terms of interaction flow and content alignment, both the transition from enactment to feedback (Q6) and the relevance to plot and character state (Q7) received support, addressing the needs captured in DG2. For transition, 13 out of 14 participants selected positive scores. For relevance, 10 out of 14 participants selected positive scores. These numbers indicate that participants generally found the interaction design natural and the feedback consistently relevant. Smooth transitions and contextually aligned feedback help sustain a coherent reasoning flow, which allows screenwriters to reflect without being disrupted by breaks in logic or mismatches in narrative context.} P3 observed \textit{``The questions raised were indeed closely aligned with the current plot development.''} Regarding the transition from inner thoughts to questions, P3 also said \textit{``The interaction from enactment of inner thoughts to raising questions was quite smooth.''}  P4 also noted \textit{``I think the questions it raised and the inner thoughts it presented earlier, overall, look quite smooth...the transition is coherent.''} \revise{On the consistency between inner thoughts and the user's imagined character psychology (Q3), overall evaluations were more varied. For imagination alignment, 6 out of 14 participants selected positive scores.}  Some users found the enactments highly consistent with their own expectations. P1 said \textit{``It was consistent with how I imagined the character's inner thoughts.''} In other cases, participants noted small mismatches when characters had unusual or multi-layered identities. P9 mentioned \textit{``For the character Hirose, it was completely off, because when analyzing the inner thoughts, it ignored that in this scene the character actually had a special fake identity.''} Interestingly, some participants appreciated such deviations. P4 said the AI added a \textit{``domineering feeling''} to the character, which, though inconsistent with her original design, \textit{``made the character more three-dimensional… quite interesting.''} This shows that divergence from user expectations was not always negative and could sometimes be interpreted as creative supplementation.  

\subsubsection{\revise{DuoDrama Leads to Insight and Refinement Willingness for Reflection}}
\revise{The constructiveness of feedback (Q9) and their ability to prompt rethinking of characters or plots (Q10) received clearly positive evaluations, addressing the needs captured in DG3. For constructiveness, 11 out of 14 participants selected positive scores. For rethinking, 9 out of 14 participants selected positive scores. These results indicate that the system effectively guided user reflection and encouraged refinement intentions.} Many participants highlighted the value of such feedback. P4 emphasized \textit{``I rated Q10 the highest score because the AI raised questions I had never considered… it prompted me to re-examine the direction of characters and plot development.''} P12 described the feedback as \textit{``extremely valuable''} and \textit{``very constructive,''} even noting they were \textit{``so precise that I never expected anyone could ask so accurately.''} P9 also saw greater \textit{``potential for exploration''} in the feedback. By contrast, ratings of whether inner thoughts provided new inspiration (Q4) were more mixed. \revise{For inspiration, 7 out of 14 participants selected positive scores.}  Some participants gained fresh perspectives. P5 said the AI enriched supporting characters and added emotional depth to the protagonist, which might inspire him to \textit{``revise my lines.''} P12 described experiencing a \textit{``sudden spark in the character.''} P4 mentioned it offered \textit{``unexpected new ideas.''} Others felt that while the enactments were accurate and aligned with their expectations, they did not always bring surprises. P10 noted \textit{``There was no surprise… just things I could already think of.''} P3 remarked \textit{``Not particularly inspiring, because it was quite ordinary.''} These responses suggest that while the system's enactments were generally reliable and coherent, their ability to spark novelty varied, with some participants valuing the inspiration they received and others experiencing them more as confirmations of their existing ideas.

\subsubsection{\revise{DuoDrama Provides Acceptable Timing and Quantity of Feedback}}
\revise{Evaluations of the timing (Q11), attunement (Q12), and quantity (Q13) of feedback were more evenly distributed, reflecting diverse user needs, addressing the needs captured in DG4. For timing, 7 out of 14 participants selected positive scores. For attunement and quantity, 8 out of 14 selected positive scores.} Some participants felt that the timing and amount of feedback were appropriate and did not disrupt their creative flow. P5 said \textit{``The timing and quantity were both accurate… just right.''} In contrast, others experienced interruptions, noting that the ``instant feedback'' were overly frequent, which disrupted their thinking. P2 said \textit{``The intervals between questions were too short, so it still interrupted me.''} At the same time, a different group of participants expressed the opposite concern, wishing for more feedback and finding the current number insufficient. P14 said \textit{``The current question number is still relatively few.''} P13 said \textit{``I even wanted it to ask more. I found the current questions quite interesting.''} Overall, these results suggest that divergent preferences point to the need for future designs to incorporate adaptive mechanisms that can dynamically adjust feedback timing and density to suit each user’s creative rhythm and emotional state.

\begin{table*}[]
    \centering
    \caption{Results of Wilcoxon Signed-Rank Tests Comparing \revise{\textit{DuoDrama}} with Other Conditions in Session 2. The conditions compared are \condone : \textit{Evaluation role} perspective with \textit{experience role}'s PE; \condtwo : \textit{Experience role} perspective with PE; \condthree : \textit{Evaluation role} perspective without PE; and \condfour : Screenplay reviewer perspective without PE.}
    \label{tab:wilcoxon_results}
    \footnotesize
    \renewcommand{\arraystretch}{1.8}
    \setlength{\tabcolsep}{2.5pt}

    \begin{tabular}{@{}l@{\hskip 1em}p{2.5cm} c cccc cccc cccc@{}}
        \toprule
        & & \textbf{\textit{DuoDrama}} & \multicolumn{4}{c}{\textbf{\revise{DuoDrama} vs. \condtwo{}}} & \multicolumn{4}{c}{\textbf{\revise{DuoDrama} vs. \condthree{}}} & \multicolumn{4}{c}{\textbf{\revise{DuoDrama} vs. \condfour{}}} \\
        \cmidrule(lr){3-3} \cmidrule(lr){4-7} \cmidrule(lr){8-11} \cmidrule(lr){12-15}
        \textbf{Dimension} & \textbf{Subdimension} & M (\revise{DuoDrama}) & M (\condtwo{}) & $W$ & $p$ & $r$ & M (\condthree{}) & $W$ & $p$ & $r$ & M (\condfour{}) & $W$ & $p$ & $r$ \\
        \midrule
        
        \multirow{5}{*}{Alignment (DG2)} & Character emotion & 5.0 & 5.0 & 43.5 & .634 & 0.13 & 4.5 & 26.5 & .231 & 0.32 & 3.5 & 16.0 & \textbf{.044*} & 0.54 \\
        & Behavioral motivation & 6.0 & 5.0 & 26.5 & .180 & 0.36 & 5.0 & 20.0 & .146 & 0.39 & 3.0 & 7.0 & \textbf{.014*} & 0.66 \\
        & Character relationship & 5.0 & 5.0 & 33.0 & .437 & 0.21 & 4.0 & 13.5 & \textbf{.041*} & 0.55 & 4.0 & 9.0 & \textbf{.018*} & 0.63 \\
        & Plot pacing & 5.0 & 3.5 & 29.0 & .238 & 0.32 & 3.0 & 7.0 & \textbf{.006**} & 0.73 & 4.0 & 10.0 & .053 & 0.52 \\
        & Thematic consistency & 6.0 & 5.0 & 33.0 & .295 & 0.28 & 4.5 & 20.0 & .078 & 0.47 & 5.0 & 19.0 & .129 & 0.41 \\
        \midrule

        \multirow{5}{*}{Quality (DG1)} & Content richness & 6.0 & 5.0 & 32.0 & .199 & 0.34 & 3.0 & 11.0 & \textbf{.008**} & 0.70 & 4.0 & 10.0 & \textbf{.016*} & 0.64 \\
        & Clarity & 5.0 & 5.0 & 22.0 & .187 & 0.35 & 4.5 & 9.5 & \textbf{.031*} & 0.58 & 6.0 & 41.5 & .610 & 0.14 \\
        & Relevance & 6.0 & 6.0 & 31.0 & .611 & 0.14 & 5.5 & 10.0 & .085 & 0.46 & 4.5 & 0.0 & \textbf{.009**} & 0.70 \\
        & Comprehensibility & 6.5 & 6.0 & 30.0 & .563 & 0.15 & 5.0 & 19.0 & .186 & 0.35 & 6.0 & 28.0 & .356 & 0.25 \\
        & Specificity of detail & 6.0 & 5.0 & 30.0 & .215 & 0.33 & 3.5 & 5.0 & \textbf{.005**} & 0.76 & 3.0 & 3.0 & \textbf{.002**} & 0.84 \\
        \midrule
        
        \multirow{6}{*}{Perceived Effectiveness (DG3)} & Emotional insight & 6.0 & 4.0 & 25.5 & .160 & 0.38 & 4.0 & 7.5 & \textbf{.023*} & 0.61 & 4.0 & 7.0 & \textbf{.009**} & 0.69 \\
        & Motivational insight & 6.0 & 5.0 & 19.0 & .130 & 0.40 & 4.5 & 14.0 & \textbf{.032*} & 0.57 & 5.0 & 25.5 & .126 & 0.41 \\
        & Relationship insight & 6.0 & 4.0 & 27.0 & .191 & 0.35 & 4.0 & 7.5 & \textbf{.010**} & 0.69 & 3.5 & 0.0 & \textbf{.004**} & 0.78 \\
        & Plot pacing insight & 5.0 & 3.5 & 22.5 & .087 & 0.46 & 3.0 & 18.0 & .080 & 0.47 & 5.0 & 36.0 & .559 & 0.16 \\
        & Thematic insight & 5.0 & 2.0 & 16.5 & .065 & 0.49 & 3.0 & 21.5 & .250 & 0.31 & 4.0 & 29.0 & .519 & 0.17 \\
        & Revision motivation & 6.0 & 2.5 & 0.0 & \textbf{.004**} & 0.78 & 4.0 & 0.0 & \textbf{.004**} & 0.78 & 6.0 & 10.0 & .085 & 0.46 \\
        \midrule

        \multirow{2}{*}{User Reflection (DG3)} & Depth of reflection & 6.0 & 4.5 & 7.0 & \textbf{.009**} & 0.69 & 5.0 & 8.5 & \textbf{.017*} & 0.64 & 5.5 & 20.0 & .308 & 0.27 \\
        & Richness of reflection & 5.5 & 3.5 & 23.0 & .076 & 0.47 & 2.0 & 0.0 & \textbf{.002**} & 0.84 & 3.5 & 13.5 & \textbf{.015*} & 0.65 \\
        \bottomrule
    \end{tabular}
    
    \par
    \smallskip
    \parbox{\textwidth}{\footnotesize Test statistics from Wilcoxon signed-rank tests include the statistic ($W$), significance level ($p$), effect size ($r$) and median (M).

    * $p < .05$, ** $p < .01$, *** $p < .001$.}
\end{table*}

\subsection{Study Session 2 Results: Offline Comparative Evaluation}

We conducted an offline comparative evaluation across four conditions for further validation. Results show that \condone exhibited significant advantages across multiple dimensions of performance, which can be further illustrated by comparisons with the other three conditions (refer to Table~\ref{tab:wilcoxon_results}).

\subsubsection{\revise{DuoDrama Provides Higher Quality Feedback for Reflection}}
In the quality dimensions, many subdimensions indicate that \condone significantly outperformed \condthree and \condfour, \revise{addressing the needs captured in DG1}. Specifically, compared to \condthree, \condone yielded significantly higher \textit{content richness} (p=.008), \textit{clarity} (p=.031), and \textit{specificity} (p=.005). Compared to \condfour, \condone showed significant advantages in \textit{content richness} (p=.016), \textit{relevance} (p=.009), and \textit{specificity} (p=.002). \revise{Higher richness, clarity, and specificity give screenwriters concrete and interpretable cues about where and why a moment may require refinement, which strengthens their reflection.}

Feedback without PE was criticized for being overly abstract. P2 described \condthree and \condfour as providing feedback that was \textit{``relatively broad, without elaborating on the specific problems''}, and P13 noted that the feedback was basic, addressing issues already resolved in early development. P11 characterized such feedback as \textit{``diagnosis at a general level''}, lacking depth and detail.  

By contrast, \condone, grounded in PE, produced more concrete and well-evidenced reflections. P1 praised its content richness, noting that it was \textit{``very divergent... it includes every point that could possibly be considered''}. P14 echoed this view, remarking that \condone's content was \textit{``richer and also more specific''}. P4 further highlighted that its \textit{``number of questions is much higher than in \condtwo{}, \condthree{}, and \condfour{}. And it's more detailed... it can really reflect the very detailed shifts in their emotions''} (refer to P4's case in Table~\ref{tab:case_full}).

\begin{table*}[t]
  \centering
  \footnotesize
  \caption{Participant Cases on Alignment and Quality. The conditions compared are \condone : \textit{Evaluation role} perspective with \textit{experience role}'s PE; \condtwo: \textit{Experience role} perspective with PE; \condthree: \textit{Evaluation role} perspective without PE; and \condfour: Screenplay reviewer perspective without PE.}
  \label{tab:case_full}
  \begin{tabularx}{\textwidth}{l l >{\RaggedRight}X >{\RaggedRight}X >{\RaggedRight}X >{\RaggedRight}X}
    \toprule
    \textbf{Dimension} & \textbf{ID} & \textbf{\condone} & \textbf{\condtwo} & \textbf{\condthree} & \textbf{\condfour} \\
    \midrule
    
    Alignment (DG2) & P1 & 
    \textbf{Question:} At this point, Lin Xi spoke quite naturally about her vulnerability and failures. But couldn't she have conveyed more of her inner resistance to going home through a bit more body language, such as her gaze at the tiny flying insects on the lamp in the market? \newline\textbf{Reason:} In this part of the scene, she narrates her thoughts as she watches the small flying insects hitting the lights and projecting her own struggles onto them. This imagery actually has a latent resonance with her unwillingness to go home and her habitual failures. Using subtle physical details will allow the emotions to flow more naturally and avoid isolated verbal expressions.&
    \textbf{Question:} At this moment, I desperately want to be acknowledged. Is it too easy for me to give myself up for just one word of affirmation from others? \newline \textbf{Reason:} I am always sensitive and agitated due to a lack of acknowledgment. I fear my emotions being controlled by external factors, which ultimately leads to more disappointment. &
    \textbf{Question:} Regarding Lin Xi's decision not to return home, is it driven more by self-esteem or the deep-seated estrangement from her mother? \newline \textbf{Reason:} She lies to her mother about finding a formal job and evades the invitation to return. We suggest the scriptwriter add related plot points to strengthen her true motivation. &
    \textbf{Question:} Does the scene's pacing feel repetitive or drawn out across the two segments—the market and the phone call at the gallery? \newline \textbf{Reason:} Lin Xi's two lengthy phone calls, one at the market and one at the gallery, make the pacing feel somewhat loose. Condensing these scenes could help focus and advance the main plot more effectively. \\
    
    \midrule
    
    Quality (DG1) & P4 &
    \textbf{Question:} I feel that when Chen Mo names his new work at the end, his acceptance of himself and his past seems too abrupt, causing a disconnect for the audience following his emotional journey. \newline \textbf{Reason:} One moment, Chen Mo is on the verge of pain and collapse; the next, he decisively tears up the resignation letter to write a new book. While this shows determination, the emotional transition lacks sufficient buildup from his years of torment and doubt. The act of tearing the letter and tossing the scraps to the sky feels rushed; a more detailed psychological portrayal would make the transformation more convincing. &
    \textbf{Question:} How can I suddenly let go of a decade of resentment and pain? Isn't that a bit unrealistic? \newline \textbf{Reason:} Every step before this was a struggle with self-doubt and guilt, yet he is suddenly able to move on. This lacks the pull of a gradual emotional shift, making the resolution feel less smooth and substantial. &
    \textbf{Question:} When Chen Mo decides to write the new book *"Nucleus and Image,"* is his emotional state one of hope or complete determination to move on? \newline \textbf{Reason:} He faces the "end of the story" and also "sets off again" in parallel. The old journey is over, but a new one begins, which could provide the actors with more spiritual states to portray. &
    \textbf{Question:} Does the ending feel too rushed, impacting the emotional progression? \newline \textbf{Reason:} The ending resolves in a series of quick actions—tearing the letter, writing the new book, pressing game buttons. This rapid pace makes it difficult for the audience to fully absorb Chen Mo's release and rebirth. \\
    \bottomrule
  \end{tabularx}
  
\end{table*}

\begin{table*}[t]
\ContinuedFloat
  \centering
  \caption{Participant Cases on Perceived Effect and User Reflection (Continued). The conditions compared are \condone: \textit{Evaluation role} perspective with \textit{experience role}'s PE; \condtwo: \textit{Experience role} perspective with PE; \condthree: \textit{Evaluation role} perspective without PE; and \condfour: Screenplay reviewer perspective without PE.}
  \footnotesize
  \label{}
  \begin{tabularx}{\textwidth}{l l >{\RaggedRight}X >{\RaggedRight}X >{\RaggedRight}X >{\RaggedRight}X}
    \toprule
    \textbf{Dimension} & \textbf{ID} & \textbf{\condone} & \textbf{\condtwo} & \textbf{\condthree} & \textbf{\condfour} \\
    \midrule
    
    Perceived Effectiveness (DG3) & P14 &
    \textbf{Question:} Su Xia holds a central position in her group of friends, but her actions often seem too passive. Is her narrative arc primarily driven by the choices of others? \newline \textbf{Reason:} For example, on the rooftop, she silently agrees with Zhang Xi's suggestion not to choose Shen Ran. This setup naturally fits her gentle personality, but the entire story consists of passive reactions, which can flatten the character arc. We suggest giving her more agency to make pivotal decisions or take action at key moments. &
    \textbf{Question:} Do I really need to get involved and sift through the messy relationships of this group? \newline \textbf{Reason:} If I don't care about this clique, it will be clear to my friends that my relationship with Lin Xi is different, affecting my image within our social circle. &
    \textbf{Question:} Are Su Xia's relationships with the other girls superficial rather than deeply connected? \newline \textbf{Reason:} Su Xia is very cheerful at night with her sisters, but matters concerning Lin Xi are lightly skimmed over. Su Xia almost seems to live in two different worlds. We suggest focusing on the emotional conflicts she experiences. &
    \textbf{Question:} Is Su Xia's emotional reaction after the breakup portrayed too lightly? \newline \textbf{Reason:} The breakup is a key plot point, but she only seems absent-minded for a moment, and her deep-seated emotional turmoil is not explored. This lessens the audience's empathy and connection with the character. \\
    
    \midrule

    User Reflection (DG3) & P12 &
    \textbf{Question:} The soldier says "Keep waiting, maybe tomorrow will be different," but shouldn't the emotional tone be more helpless and bleak, rather than comforting the youth? \newline \textbf{Reason:} The atmosphere of entrapment is constant, from the suffocating daily life to the civil unrest. The soldier's character arc is also defined by this. Adding too much hope here could contradict the established themes of "sometimes there's time, sometimes there isn't," and "waiting for so long it's unclear," weakening the thematic consistency. &
    \textbf{Question:} I say "Keep waiting, maybe tomorrow will be different," but in my heart, I don't really believe tomorrow will change. Isn't this kind of dialogue just empty talk? \newline \textbf{Reason:} This dialogue, filled with a sense of hope, clashes with my established character of being weary and powerless. It adds unnecessary complexity and makes it hard for the audience to understand what I'm really thinking. &
    \textbf{Question:} The dialogue between Soldier A and Soldier B about "waiting for time" seems a bit repetitive. Are they essentially the same? \newline \textbf{Reason:} The two characters' dialogues on waiting both lack force. For example, "We just have to keep waiting," "...we're used to it anyway." We confirm that these are intended to build a shared foundation, but we suggest enriching the emotional texture. &
    \textbf{Question:} The theme is expressed too obscurely, making it difficult for the audience to grasp the core conflict. \newline \textbf{Reason:} The themes of "fermentation" and "waiting" are repeated as key symbols, but without clear points of resolution, it may lead the audience to lose sight of the world's rules and the character's objectives, making it difficult to understand the intended meaning. \\
    \bottomrule
  \end{tabularx}
\end{table*}

\subsubsection{\revise{DuoDrama Shows Stronger Aligned Feedback for Reflection}}
In the alignment dimension, \condone significantly outperformed both \condthree and \condfour, \revise{addressing the needs captured in DG2}. Compared to \condthree, \condone achieved higher alignment in \textit{character relationships} (p=.041) and \textit{narrative pacing} (p=.006). Compared to \condfour, \condone was superior in \textit{character emotions} (p=.044), \textit{behavioral motivation} (p=.014), and \textit{character relationships} (p=.018). \revise{When feedback aligns closely with the character’s emotional, motivational, and relational logic, it helps screenwriters reflect more effectively because they can focus on the issues rather than resolving misunderstandings or correcting misinterpretations from AI feedback.}

These results suggest that feedback without PE was more prone to drift away from the core content and creative intentions of the screenplay. P9 criticized \condfour's feedback as \textit{``not really addressing the screenplay issues''} and even \textit{``questioning the theme itself,''} which he regarded as fundamentally misaligned with the refinement goal. Similarly, P5 observed that \condthree's feedback focused on a \textit{``rather marginal tool or character,''} which he found \textit{``irrelevant and ineffective,''} reflecting a misjudgment of narrative importance. P1 further noted that \condfour treated screenplays as \textit{``traditional text,''} overlooking the essential role of audiovisual language.  

By contrast, \condone demonstrated a professional and comprehensive understanding that participants considered highly aligned. P14 lauded its ability to accurately capture \textit{``the entire screenplay, the girl's personal growth arc, including her internal changes,''} reflecting a holistic grasp of the narrative. P1 similarly emphasized that \condone's perception was \textit{``very on point,''} analyzing even \textit{``capturing the character's inner emotions and every concrete action in such detail''} and thereby maintaining strong alignment with the character's emotions and behavioral motivation in the screenplay (see P1's case in Table~\ref{tab:case_full}).

\subsubsection{\revise{DuoDrama Enhances the Perceived Effectiveness, Depth, and Richness of Human Reflection}}
In many subdimensions of perceived effectiveness and user reflection, \condone significantly outperformed all other conditions, \revise{addressing the needs captured in DG3}.  

First, \condone was significantly stronger than \condtwo in \textit{revision motivation} ($p=.004$) and \textit{depth of reflection} ($p=.009$). Participants felt \condtwo's perspective was overly immersive and lacked critical distance. P9 described it as \textit{``too immersed in the screenplay''}, while P14 remarked that it resembled \textit{``a reproduction of the character's inner thoughts''} without external critical analysis, offering little motivation to revise. In contrast, \condone, by adopting a challenging yet professional stance, was viewed as a powerful catalyst for change. P1 described it as providing the \textit{``most inspiration''}, emphasizing that it offered a unique and professional analysis that \textit{``covered everything from themes to small props and actions''}, giving him unexpected entry points and fresh ideas. P14 likewise felt \condone spoke \textit{``like an industry peer,''} offering a detached yet constructive third-party \textit{evaluation role}'s perspective that exposed blind spots and provided insights grounded in both micro-level details and macro-level growth arcs (see P14's case in Table~\ref{tab:case_full}). This combination of breadth and precision was repeatedly described as stimulating genuine motivation for refinement.  

Second, \condone significantly outperformed \condthree in \textit{emotional insight} ($p=.023$), \textit{motivational insight} ($p=.032$), \textit{relationship insight} ($p=.010$), \textit{revision motivation} ($p=.004$), \textit{depth of reflection} ($p=.017$), and \textit{richness of reflection} ($p=.002$). Participants considered \condthree's feedback superficial and disconnected from the experiential grounding needed to reveal deeper dynamics. P3 noted that without experiential grounding, the feedback remained \textit{``generic and shallow''}. By contrast, \condone's feedback not only directed attention to how emotions and motivations shaped character relationships but also offered concrete and actionable refinement paths. For example, P14 suggested adding \textit{``a subconscious hand gesture''}, providing participants with clear starting points to revise. P14 emphasized that such concrete yet challenging feedback made them \textit{``want to revise the screenplay the most''}, because it dared to raise uncomfortable but constructive critiques, while P4 and P14 both noted that it revealed \textit{``suppressed emotions or hidden gaps''}. This blend of constructive \textit{``offense''} and clarity was perceived as energizing rather than discouraging, driving strong revision motivation.  

Third, \condone also significantly outperformed \condfour in \textit{emotional insight} ($p=.009$), \textit{relationship insight} ($p=.004$), and \textit{richness of reflection} ($p=.015$). Participants criticized \condfour's feedback as overly abstract and detached. P13 remarked that it \textit{``felt like from our teachers,''} with feedback that was \textit{``too high-level and abstract''}. P9 similarly likened it to \textit{``a clueless client asking irrelevant multi-dimensional questions''}. In contrast, \condone's experience-grounded feedback explicitly addressed emotional and relational dynamics and prompted deeper and more systemic reflection. P12 provided a clear example, explaining that one \condone prompt triggered a powerful realization: \textit{``After reading it, I was indeed reflecting on the entire screenplay, feeling that the earlier character emotions were too suppressed for the whole screenplay''} (see P12's case in Table~\ref{tab:case_full}). Others echoed this, with P14 noting that \condone's feedback forced them to \textit{``connect local moments with overall structure''} and P12 adding that they prompted screenwriters to \textit{``revisit the core themes and intentions of their writing''}. By pushing participants to reconsider why at the level of concept and theme, \condone's feedback went beyond local fixes to stimulate the richness of reflection.   

\subsection{Summary of \textit{DuoDrama} Advantages}  
\revise{Overall, participants emphasized three advantages of \textit{DuoDrama}. First, the system raises feedback that was rich, detailed, clear, and relevant, extending from core emotional motivations to subtle prop-level cues and fine-grained actions. Second, its feedback demonstrated a professional and well-aligned grasp of the screenplay, connecting local enactments, including character emotion and motivation, with broader narrative structures, including character relationships and plot pacing. Third, this combination of high-quality and contextually aligned feedback stimulated reflection and motivation for refinement. Participants noted that \textit{DuoDrama} revealed blind spots, surfaced overlooked dynamics, and offered constructive challenges that inspired new directions for refinement. Together, these qualities made \textit{DuoDrama} the most consistently valued condition and positioned it as a strong form of support for screenwriters' reflection in refinement.}



\section{Discussion}


\revise{Our evaluation shows that by providing high-quality, contextually aligned, and well-timed feedback, \textit{DuoDrama} enhances screenwriters’ reflection with more effectiveness, depth, and richness. We conclude by discussing broader implications for designing future AI-assisted reflection in human–AI collaboration.}

\subsection{\revise{Experience-Grounded Feedback in Supporting Human Reflection}}

\revise{Our results show that \textit{DuoDrama} strengthens screenwriters’ reflection by enabling the agent to draw on its enacted personal experience when generating feedback. This experiential grounding produces feedback that is richer, more detailed, and better aligned with the specific context. Screenwriters reported that, by making the experiential grounding of each feedback explicit, such as why a character could or could not take an action or how a moment feels from within the scene, they could more easily reflect, trace insights back to concrete contexts, and determine whether refinement was needed.} \revise{At the same time, our study shows that effective support for human reflection requires balancing immersion and distance in feedback for refinement. When feedback draws only from the character perspective, it may align too closely with character logic and overlook broader narrative considerations. When feedback lacks experiential grounding, it often becomes generic and less actionable for refinement. \textit{DuoDrama} combines these two perspectives: experiential grounding offers realism and specificity grounded in internal experience, while evaluative distance enables a broader narrative assessment. Together, we strengthen reflection by helping screenwriters identify overlooked issues and examine their decisions from both perspectives, avoiding the narrowness of full immersion and the limitations of detached critique.}

\revise{Looking ahead, one possible extension is to enrich human reflection through more differentiated perspectives. This may include instantiating specific personas, experimenting with varied configurations of \textit{experience roles} and \textit{evaluation roles}, or exploring cross-scenario reflection, where experiential perspectives from one context are used to interrogate and enrich another. More broadly, such designs build on the principle of perspective-taking: shifting between differentiated perspectives can surface tensions, alternatives, and overlooked assumptions that a single perspective might miss~\cite{10.1145/3719160.3736634, 10.1145/3544548.3580763}. These extensions could broaden the range of triggers for human reflection, equipping people with awareness of alternative interpretations and strategies~\cite{10.1145/3613904.3642530}.}

\revise{Overall, by producing feedback that is traceable to experience, \textit{DuoDrama} helps bridge the gap between assessment and actionable guidance. This balance between internal immersion and external distance is not limited to screenwriting refinement. It reflects a broader need in domains where feedback should connect situated experience with higher-level evaluation. This suggests that the underlying workflow, \textit{ExReflect}, also holds potential for generalizability across domains. For example, educational feedback often requires understanding how a learner arrives at a misconception while also offering structured guidance for correction~\cite{lim2025feed, 10.1145/3610591.3616427, 10.1145/3610537.3622957}. A system that alternates between a learner-like experience role and an instructor-like evaluation role could support this requirement. Likewise, in design work, practitioners move between imagining how users encounter an interface and evaluating the design from a professional perspective~\cite{lowgren2007thoughtful, stone2005user, 10.1145/3678884.3681890}. Providing feedback from both perspectives can clarify why an interaction causes confusion and how it might be improved. Thus, by operationalizing two-perspective reflection, \textit{ExReflect} also shows potential for supporting reflection in other fields where coordinating contextual experience understanding with evaluative perspective is essential~\cite{10.1145/3173574.3174223, 10.1145/3290605.3300526}.}

\subsection{Adaptive Regulation: Timing, Quantity, and Personalization}
Participants reported variation in their preferences regarding both the timing and the quantity of feedback. This variation suggests that future AI systems designed to support human reflection may benefit from adaptive regulation mechanisms, rather than relying on one-size-fits-all strategies~\cite{796083, chen2025we}. Taken together, these findings point to design considerations for future AI-assisted human reflection.

First, regarding timing, future systems may flexibly alternate between \textit{instant feedback} and \textit{post-hoc feedback}, while also supporting opportunities for extended conversational follow-ups. As P1 noted, feedback in current systems often lacked chances for continued dialogue, which disrupted the flow of human–AI co-creation. To address this issue, systems could incorporate dialogic follow-up mechanisms and become stage-aware, offering broader thematic or structural prompts during early exploration and more fine-grained, line-by-line guidance during late-stage refinement. In addition, as our study showed a general preference for \textit{post-hoc feedback} (see Sec.~\ref{sec:Session 1}), future work could further examine how different timing strategies shape human reflection~\cite{10.1145/3635636.3656183, kreminski2024intent}.

Second, regarding quantity, participants expressed diverging preferences. Some perceived frequent \textit{instant feedback} as overwhelming and disruptive, while others preferred denser guidance. This divergence suggests that future systems could allow users to set, or dynamically adjust, the frequency and number of feedback items, helping balance reflective support with creative flow. In broader multi-agent settings, such regulation may also extend to how reflective contributions are distributed across agents. For example, systems could monitor which characters carry the narrative load within a scene and prioritize more feedback from those characters’ perspectives, while limiting other, more peripheral roles to essential clarifications.

Overall, incorporating adaptive regulation mechanisms across timing, quantity, and personalization may help AI-assisted human reflection systems better align with individual users’ creative practices and preferences~\cite{10.1145/3706598.3713883, orzikulova2024time2stop}.

\subsection{Limitations and Future Work}

\revise{While our study demonstrates the effectiveness of \textit{DuoDrama}, it also has limitations. First, the work was situated in screenwriting, which we chose as a rich testbed for AI-assisted human reflection. As a result, the study focuses on a single creative practice, calling for future exploration in broader domains. Second, the evaluation was short-term. Although participants used their own screenplays, each engaged in only two sessions, leaving their longer-term influence on human–AI collaboration to be examined in future work.} Future research can extend our study in several directions. Longitudinal deployments are needed to understand how repeated interaction with \textit{DuoDrama} shapes users’ reflection and refinement practices over time. Besides, beyond our current focus on text-based personal experience, future work may explore grounding feedback in multi-modal forms of experience. For instance, experiential evidence could be derived not only from dialogue but also from visual renderings or auditory cues such as tone and pacing. Such multi-modal grounding would allow agents to capture a richer spectrum of experience and address aspects that extend beyond text alone. \revise{In addition, this study focused on general refinement after a completed draft. Future work may extend support to more specific refinement stages, such as first-draft or late-stage polishing, which may produce different outcomes. Future systems could also incorporate other reasoning paradigms, such as analogical or counterfactual reasoning, to offer more diverse and nuanced forms of AI feedback. Another direction is to explore adaptive orchestration in multi-agent settings, where agents with distinct roles and experiences communicate with one another before presenting feedback to users. Together, these point to opportunities for expanding \textit{DuoDrama} and advancing its implications across a broader range of human–AI collaboration.}


\section{Conclusion}

\revise{This study introduces \textit{DuoDrama}, a system that enables AI to generate feedback coordinated across two perspectives to support screenwriters’ reflection in refinement. To enable \textit{DuoDrama}, we design the \textit{ExReflect} based on the insights from performance theories and formative study. \textit{ExReflect} coordinates between two roles: an \textit{experience role}, where the agent performs immersive enactment to produce a character’s inner thoughts as personal experience, and an \textit{evaluation role}, where the agent shifts to the actor portraying that character to provide feedback informed by this experience. \textit{DuoDrama} employs \textit{ExReflect} within a multi-agent architecture due to the multi-character nature of the screenplay. A user study with fourteen professional screenwriters shows that \textit{DuoDrama} improves the quality and alignment of feedback and enhances the perceived effectiveness, depth, and richness of reflection. Looking forward, \textit{DuoDrama} offers a direction for designing AI-assisted reflection that balances internal experience with external evaluation. Future opportunities include extending the implications of \textit{DuoDrama} and \textit{ExReflect} to support human reflection in other areas of human–AI collaboration. This extension may be advanced through directions like expanding experiential grounding into multi-modal forms, developing adaptive multi-agent orchestration, and investigating alternative reasoning paradigms.}

\begin{acks}
This work was partially supported by the Research Grants Council of the Hong Kong Special Administrative Region under the General Research Fund (GRF) (Grant No. 16218724 and No. 16207923) and is part of the AFMR collaboration supported by Microsoft Research. We thank Jiaxiong Hu, Runhua Zhang, Baiqiao Zhang, Linping Yuan, and Shuchang Xu for their valuable discussions and support on this work. We also sincerely thank all participants for their time and contributions. Finally, we greatly appreciate the reviewers for their insightful and constructive feedback.
\end{acks}


\bibliographystyle{ACM-Reference-Format}  
\bibliography{software}  




\end{document}